\documentclass[10pt,journal,compsoc]{IEEEtran}
\usepackage{amsthm,amsmath,amssymb}
\usepackage{lipsum}
\usepackage{cuted}
\usepackage{mathrsfs}
\usepackage{mathtools}
\usepackage{algorithm}
\usepackage{algorithmic}
\usepackage{amsfonts}
\usepackage{tabularx}
\usepackage{booktabs}
\usepackage{ifpdf}
\usepackage[justification=centering]{caption} 
\usepackage{pifont}

\usepackage{diagbox}
\usepackage{caption}
\usepackage{subfigure}
\usepackage{graphicx}
\usepackage{threeparttable}
\usepackage{multicol}
\usepackage{verbatim}
\usepackage{multirow}

\usepackage{bigstrut}
\usepackage{hyperref}
\usepackage[table]{xcolor}
\usepackage{makecell}

\ifCLASSOPTIONcompsoc
  \usepackage[nocompress]{cite}
\else
  \usepackage{cite}
\fi
\ifCLASSINFOpdf
\else
\fi
\begin{document}
%
\title{PBAG: A Privacy-Preserving Blockchain-based Authentication Protocol with Global-updated Commitment in IoV}
%
%
%
%

\author{Xia~Feng,
        Kaiping~Cui
        and~Liangmin~Wang,~\IEEEmembership{Member,~IEEE},
\IEEEcompsocitemizethanks{\IEEEcompsocthanksitem X. Feng and K. Cui are with the School of Automotive and Traffic Engineering, Jiangsu University, Zhenjiang,
212013, China (e-mail: xiazio@ujs.edu.cn).
\IEEEcompsocthanksitem L. Wang are with School of Cyber Science and Technology, Southeast University, Nanjing, 211110, China (e-mail: liangmin@seu.edu.cn).}
\thanks{\textit{Corresponding author: Xia Feng}.}}

%
%

\markboth{Journal of \LaTeX\ Class Files,~Vol.~14, No.~8, August~2015}%
{Shell \MakeLowercase{\textit{et al.}}: Bare Demo of IEEEtran.cls for Computer Society Journals}
%



\IEEEtitleabstractindextext{%
\begin{abstract}
Internet of Vehicles(IoV) is increasingly used as a medium to propagate critical information via establishing connections between entities such as vehicles and infrastructures. During message transmission, privacy-preserving authentication is considered as the first line of defence against attackers and malicious information. To achieve a more secure and stable communication environment, ever-increasing numbers of blockchain-based authentication schemes are proposed. At first glance, existing approaches provide robust architectures and achieve transparent authentication. However, in these schemes, verifiers must connect to the blockchain network in advance and accomplish the authentication with smart contracts, which prolongs the latency. To remedy this limit, we propose a privacy-preserving blockchain-based authentication protocol(PBAG), where Root Authority(RA) generates a unique evaluation proof corresponding to the issued certificate for each authorized vehicle. Meanwhile, RA broadcasts a public global commitment based on all valid certificates. Instead of querying certificates stored in the blockchain, the vehicle will be efficiently proved to be an authorized user by utilizing the global commitment through bilinear pairing. Moreover, our scheme can prevent vehicles equipped with invalid certificates from accomplishing the authentication,  thus avoiding the time-consuming for checking Certificate Revocation List (CRL). Finally, our scheme provides privacy properties such as anonymity and unlinkability. It allows anonymous authentication based on evaluation proofs and achieves traceability of identity in the event of a dispute. The simulation demonstrates that the average time of verification is 0.36ms under the batch-enabled mechanism, outperforming existing schemes by at least 63.7$\%$.
\end{abstract}

\begin{IEEEkeywords}
 Blockchain, Authentication, Privacy-preserving, Commitment
\end{IEEEkeywords}}

\maketitle

\IEEEdisplaynontitleabstractindextext

%
\IEEEpeerreviewmaketitle

\IEEEraisesectionheading{\section{Introduction}\label{sec:introduction}}

\IEEEPARstart{T}{he} Internet of Vehicles (IoV) is the concept of connecting vehicles together in the public network, intending to predominantly facilitate the dissemination of critical information in real-time  in the Intelligent Transportation System (ITS)\cite{matsumoto2017ikp}. IoV has tremendous potential to improve road safety and traffic efficiency\cite{liu2021pptm}, especially in various applications such as congestion control, traffic management, and collision avoidance. The operation of IoV in the V2X mode enables vehicles to exchange information such as speed, location, and heading through the dedicated short-range communication (DSRC) protocol standard\cite{ma2021attribute}. Besides, the cooperative safety applications are a significant branch of vehicular networks, where the emergency message dissemination enables each vehicle to intelligently perceive surrounding conditions and timely make decisions about potential dangers\cite{liu2019tcemd}. Unfortunately, due to the openness of the wireless channel, malicious users are inevitable in the networks, which could launch arbitrary attacks. For instance, the adversaries can cause traffic accidents and traffic disturbances by sniffing, intercepting, replying, modifying or deleting traffic statutes\cite{zhou2022efficient}. Adversaries can also eavesdrop and collect beacon messages broadcast by the vehicle to determine the vehicle’s driving route and track the vehicle’s trajectory\cite{li2020papu}. Moreover, adversaries may attack the vehicle’s control system and take control of the intruded vehicles. Thus, security and privacy become crucial and indispensable issues for IoVs. In this context, reliable authentication schemes are required to effectively support information dissemination, particularly amid the transformation of the assisted driving system to self-driving vehicles.

Identification of benign vehicles forms the cryptographic basis for ensuring security and privacy in IoVs. Numerous researchers at home and abroad propose various authentication schemes, which mainly fall into three categories such as authentication schemes based on Public Key Cryptosystem \cite{adams1999understanding}, Identity-based Cryptosystem \cite{zhang2008efficient}\cite{wu2017efficient} and Group-based protocols \cite{liu2014message}\cite{jung2009robust}. Public Key Cryptosystem always needs an online Certificate Authority(CA), which often brings  communication time delay and involves certificate management problems(e.g., involving a cumbersome certificate revocation list). While the identity-Based Cryptosystem scheme can alleviate these problems, they still need to process the escrow issues. There is a risk of leakage of the secret key once Key Generation Center(KGC) is compromised.  Group signature-based scheme protects the true identity, but its challenges rely on the incurred computational and communication overheads. Moreover, the ability of privacy protection depends on the number of attended vehicles. It is obvious that the optimum number of vehicles joining the group cannot be guaranteed every time, particularly in rural areas with sparse vehicles. When the authentication process is achieved in a centralized manner, vehicles usually access the centralized server to download the records of the other vehicles in all these three categories. The centralized architecture may cause problems such as excessive computing burden and single point of failure of the central server.

\begin{sloppypar}
Currently, the emerging blockchain technology has received more attention in IoV. Blockchain can provide a robust and stable communication environment for authentication, effectively avoiding problems such as the single point of failure. Therefore, ever-increasing numbers of blockchain-based authentication schemes are proposed\cite{shrestha2018blockchain}\cite{lu2019blockchain}\cite{de2019energy}. Blockchain is mainly utilized to store certificates or status change records in such approaches, and vehicles/infrastructures perform the authentication process with smart contracts. However, existing approaches face two major challenges in IoV authentication. Firstly, vehicles/infrastructures must establish a connection with the blockchain network and accomplish the authentication with the assistance of blockchain, which prolongs authentication latency. Secondly, the execution of smart contracts is based on the consensus mechanism, which causes additional time overhead in the authentication process. 
\end{sloppypar}
In this paper, our work is to address the aforementioned efficiency and privacy problem of the existing authentication schemes. Major contributions of this paper are listed as follows:

\begin{itemize}
    \item \textbf{No interaction with blockchain.} The whole authentication process only need one end-to-end (E2E) wireless transmission between vehicles and verifiers, and vehicles do not need to interact with blockchain. That's benefiting from the public global commitment broadcast by the Root Authority(RA),  the vehicle will be efficiently proved to be an authorized user by utilizing it without querying certificates stored in the blockchain. 
    
    
    
    \item \textbf{No need to retrieve CRL.} Our proposal does not need to maintain additional Certificate Revocation List (CRL), and avoids the time-consuming for checking CRL in the the authentication process. That's because the verifier can prevent vehicles equipped invalid certificates from accomplishing the authentication with the assistance of global commitment generated based on the all valid certificates in the network. Besides, we also propose a commitment update mechanism to guarantee fast update of global commitments.

    

    \item \textbf{Privacy-preserving properties.} Our proposal guarantees security such as anonymity and unlinkability under the security assumptions. Precisely, the zero-knowledge proof enables authentication between users to be processed anonymously, and it can achieve traceability of identity in the event of a dispute. Besides, the verifier cannot link messages with a particular user during the authentication. Finally, we conduct security analyses and demonstrations that our scheme can resist common attacks.
    \item \textbf{Theoretical analysis and simulation.} Considering the real-time requirements of IoV communication scenarios, we conduct extensive simulation in \textit{Hyperledger Fabric v2.0.0} and \textit{ns-2.35} to evaluate our proposal in terms of authentication latency and information transmission efficiency. The performance evaluation indicates that the proposed protocol can signiﬁcantly reduce the authentication latency and message loss rate. With batch authentication, the authentication latency is reduced by more than 33.8\% comparing with\cite{yang2020delegating} \cite{wang2018privacy}\cite{feng2021p2ba}. Besides, the message loss rate of our scheme is controlled within 9.6\%, outperforming the common scheme and satisfying the communication requirements in IoV. 
 
 \end{itemize}

The remainder of this paper is organized as follows: Section 2 reviews some related works. Section 3 introduces several preliminaries and mathematical assumptions.  The framework and security model are formalized in Section 4. Section 5  presents PBAG, which is a privacy-preserving authentication scheme suited for IoV.  The analysis of the security and privacy of our model is in Section 6. We analyze the simulation results of our model and compare PBAG with the existing schemes in Section 7. Finally, we conclude our proposal and describe the future work in Section 8.

\section{Related Work}
Recently, a wide range of research works related to authentication has been proposed. Existing approaches can be classified into three categories\cite{shao2015threshold}such as Identity-based cryptography, Public key Infrastructure(PKI) and Blockchain-based schemes.
\subsection{Identity-based Authentication Structures }

\begin{sloppypar}

Identity-based authentication schemes have been improved by diminishing the possibility of identity disclosure, as in the case of the pseudonym\cite{vijayakumar2018computationally}\cite{wei2020secure} and group-based schemes\cite{azees2017eaap}\cite{zhang2019group}. Vijayakumar \textit{et al.} \cite{vijayakumar2018computationally} propose a privacy-preserving anonymous mutual and batch authentication scheme. In this scheme, each vehicle can generate anonymous certificate that facilely verified by other vehicles. Moreover, the scheme supports RSU to conduct batch authentication of multiple vehicles. Nevertheless, duo to changing pseudonyms frequently for privacy protection\cite{2016IEEE}, vehicles have to communicate with trust authority(TA), which cause high computational overhead and communication cost. Wei \textit{et al.} \cite{wei2020secure} propose a conditional privacy-preserving authentication protocol, which contains two main institutions, namely registration server(RS) and verification server(VS). RS generates the pseudonyms using homomorphic encryption, and VS is responsible for verifying the legitimacy of pseudonyms. In this scheme, the system architecture has higher security, reducing the reliance of the system on the central authority. Unfortunately, this scheme has the risk of disclosing user privacy information, and the computation overhead is still relatively high.
\end{sloppypar}

Chaum and Heyst \cite{chaum1991group} first propose the concept of group signature in 1991. Based on the group signature scheme, signatures can be generated by group members without revealing their real identity. Azees \textit{et al.} \cite{azees2017eaap} propose an EAAP protocol, which supports tracing electric vehicle or OBU in IoV  with the assistance of a conditional tracking mechanism. However, the scheme do not mention non-frameability, which cause the adversary can forge or tamper with signatures. For more effcient message signature verification, Zhang \textit{et al.} \cite{zhang2019group} propose  a batch group signature scheme. This scheme supports fast revocation check by conducting a group session key(GSK)-based revocation strategy(GSSA). GSSA has an excellent performance in terms of computation time cost, message delay and loss rate. Besides, GSSA can defend against common  attacks, such as tracking attacks, replay
attacks, impersonation attacks and sybil attacks. However, because the signatures are generated without challenge value, the trustworthiness of the sender’s message content cannot be recognized, which further cause the vehicle not being able to verify the legitimacy of the response from RSU.


\subsection{PKI-based Authentication }


PKI-based authentication approaches intend to ensure that information being forwarded in IoVs is authenticated. The earlier research reported in \cite{lu2008ecpp}\cite{callegati2009man} use CAs to issue certificates and keep track of the certificates along with proper verification. More concretely, the RSU authenticates the other vehicle's identity using its certificate. However, RSUs in IoV are prone to commit errors or can be compromised by adversaries. 
In recent years, researchers have also attempted to optimize the security and performance of PKI-based communication or authentication approaches in IoV\cite{joshi2017reliable}\cite{asghar2018scalable} \cite{tangade2018decentralized} \cite{wang2020hybrid}.
Considering the security issues of the Vehicle-to-Vehicle(V2V) architecture of ITS, Joshi \textit{et al.} \cite{joshi2017reliable} propose an efficient V2V communication approach. This scheme constructs one PKI-based mechanism adopting event-triggered messages transmission. Moreover, they also address several open issues with the assistance of machine learning for V2V communication scenario.
Asghar \textit{et al.} \cite{asghar2018scalable} propose a scalable and efficient PKI-based authentication protocol. Although this scheme constructs the linear-size  CRL and provides efficient authentication, the TA still needs to maintain a CRL for revoked vehicle certificates, which is inflexible. Specifically, the TA need to implement an in-depth search in CRL to find the certificate of the malicious vehicle sent forged traffic-related messages, which incurs extra time delay and computational cost.
Tangade \textit{et al.} \cite{tangade2018decentralized} propose a decentralized and scalable privacy-preserving authentication scheme, which achieves privacy message protection even in-vehicle identity revelation or information broadcast. However, the forward security of this scheme is not fully considered.
Wang \textit{et al.} \cite{wang2020hybrid} propose a hybrid conditional privacy-preserving authentication scheme, which combines PKI and identity-based signature. In this scheme, the vehicle accomplish authentication process using TA-assigned long-term certificates. Nevertheless, the PKI certificates cause huge communication cost during authentication process.


\subsection{Blockchain Based Authentication}
Blockchain technology\cite{nakamoto2008bitcoin} emerges as a promising solution for decentralized authentication for IoVs. 
Shrestha\cite{shrestha2018blockchain} proposed a blockchain-based authentication and revocation framework in IoVs. In this framework, vehicles obtain the ID and certificate from the CA and store them together in the blockchain. Thereby RSUs can verify the vehicle's identity instead of verifying identity with a trusted authority, which can reduce both the computation and communication cost. 
Feng \textit{et al.} \cite{feng2019bpas} propose a Blockchain-assisted privacy-preserving authentication scheme (BPAS). Constructing with consortium blockchain, this scheme achieves automatic authentication while privacy protection. In addition, the scheme achieves message credibility checking, vehicle behavior monitoring and communication history tracking. But, their scheme cannot support batch batch authentication of messages. 
Yao \textit{et al.} \cite{yao2019bla} propose a lightweight anonymous authentication mechanism to achieve cross-data center authentication with the assistance of blockchain and cryptographic functions. However, this scheme do not perfect the discussion and analysis of several security features, such as resistance to masquerading or session key compromise, and forward confidentiality. Besides, the scheme  cannot support bidirectional  authentication between service managers (SMs) and vehicles.
In order to achieve the tamper-resistant and traceability in authentication, Wang \textit{et al.} \cite{wang2020b} propose a paring-assisted V2I authentication scheme with trustworthiness scalable computation, which encrypts the combination of trustworthiness and attributes of vehicles, and stores ciphertext into the blockchain. However, bilinear pairing operation incurs high computational cost during authentication process in this scheme.
Xue \textit{et al.} \cite{xue2022distributed} propose a blockchain-based tamper-proof roaming
authentication scheme for mobile vehicular networks. The authors achieve an unforgeable billing scheme by deploying bloom filters on blockchain, which contributes to security and performance validations. Unfortunately, the complex security framework leads to the relatively high latency during the authentication process.
Son \textit{et al.} \cite{son2022design} propose a blockchain-based lightweight V2I handover authentication scheme, which integrates real-or-random logic and Burrows–Abadi–Needham logic. However, the authors do not optimize the overhead of block and consensus process.

Most of the blockchain-based authentication approaches face three major challenges. Firstly, such schemes need to involve  blockchain interaction during the authentication phase, which leads to longer time delay and huge computation cost. Secondly, lacking of flawless identity management mechanism. Especially, vehicles need to query a CRL to check the current status of the certificate during the authentication process, which incurs extra communication overhead. Apart from that, plenty of blockchain-based authentication schemes have not enough decentralization, and the single trusted authority has most of permissions. Therefore, the existing works cannot be fully adapted to the real-world scenario.

\section{Preliminaries and Mathematical Assumptions}\label{Preliminaries}
In this section, we introduce several preliminaries and mathematical assumptions involved in this paper.

\subsection{Lagrange Polynomial Interpolation}
We can find the unique polynomial $\Psi(X)$ for the given $n$ pairs $(x_i, y_i)_{i\in[0, n)}$ using $Lagrange$ $interpolation$\cite{berrut2004barycentric}, and it satisfies $\Psi(x_i) = y_i$. Specifically, the $\Psi(X)$ is computed as\begin{small} $\Psi(X) = {\sum}_{i\in[0, n)}\mathcal{L}_i(X)y_i$\end{small} in $O(n log^2 n)$ time\cite{von2013fast} and its degree $\textless n$, where\begin{small} $\mathcal{L}_i(X) = {\prod}_{j\in[0, n), j{\neq }i}\frac{X-x_j}{x_i-x_j}$\end{small}. It is crucial that the $\mathcal{L}_i(X)$ has property that $\mathcal{L}_i(X_i) = 1$ and  $\mathcal{L}_i(X_j) = 0, \forall i, j \in [0, n)$ with $j \neq i$.

\subsection{KZG Polynomial Commitments}\label{KZG Polynomial Commitments}
The KZG Polynomial Commitments is a constant-sized commitment scheme for degree $n$ polynomials $\Psi(X)$\cite{2010Constant}. KZG scheme can generate an evaluation proof for any $\Psi(x_i)$, which has been utilized in stateless cryptocurrencies\cite{tomescu2020aggregatable}. Besides, the generation of evaluation proof does not depend on the degree of the committed polynomial. In this paper, the parameter $\tau$ is $trapdoor$, and we can utilize the MPC protocol\cite{bowe2017scalable} to compute the public parameters $(g^{{\tau}^i})_{i \in [0, n)}$ that hides the parameter $\tau$.

\begin{itemize}
\item{\textbf{Committing.}} We define the polynomial $\Psi(X)$ of degree d $\leq$ n with coefficients $s_0, s_1,\ldots, s_d$ in $\mathbb{Z}_p$. Therefore, we can generate a KZG commitment to $\Psi(X)$, that is $C = {\prod_{i= 0}^d}(g^{\tau^i})^{s^i} = g^{{\sum}_{i= 0}^ds_i\tau^i} = g^{\Psi(\tau)}$, which takes $\Theta(d)$ time to commit $\Psi(X)$. In our scheme, $C$ is a global commitment(elaborated in \ref{Notations and definitions}).

\item{\textbf{Proving one evaluation.}} Taking the calculation of the evaluation proof that $\Psi(x) = y$ as an example. The KZG utilize the polynomial remainder theorem to transform the $\Psi(x) = y$, that is $\Psi(x) = y$ $\Leftrightarrow$ $\exists q(X), \Psi(X) - y = q(X)(X -x)$. Therefore, the proof for $\Psi(x) = y$ is just a KZG commitment to q(X): a single group element $\pi = g^{q(\tau)}$, which is computed $\Theta(d)$ time. The verifier takes constant time to compute if $e(C/g^y, g)$ = $e(\pi, g^{\tau}/g^x)$ $\Leftrightarrow$ $e(g^{\Psi(\tau)-y}, g)$ = $e(g^{q(\tau)}, g^{\tau-x})$ $\Leftrightarrow$ $e(g, g)^{\Psi(\tau)-y}$ = $e(g, g)^{q(\tau)(\tau-x)}$ $\Leftrightarrow$ $\Psi(\tau) - y = q(\tau)(\tau - x)$.

\item{\textbf{Proving multiple evaluation.}}\label{Multiple Evaluatio} We define a set of points $I$($|I|\leq d$) and the evaluations of $I$: $\Psi(i)_{i\in I}$.
First, the prover computes a(X) = ${\prod}_{i\in I}(X - i)$ in $\Theta(|I|log^2|I|)$ time. Second, the prover computes $\Psi(X)/a(X)$, and obtains a quotient $q(X)$ and remainder $r(X)$. After above calculation, the prover can achieve the equation relation: $\Psi(X) \ =\ a(X)\cdot q(X)+r(X)$ and the batch proof $\pi_I \ = \ g^{q(\tau)}$. The verifier need to compute the $a(X)$ and $r(X)$ to verify $\pi_I$ and $\{\Psi(i)\}_{i\in I}$ against $C$. In particular, verifier first computes $a(X)$ from $I$ and interpolates $r(X)$ that satisfies $r(i)=\Psi(i)$, $\forall i\in I$ in $\Theta(|I|log^2|I|)$. Second, the verifier computes $g^{a(\tau)}$ and $g^{r(\tau)}$ and checks if $e(C/g^{r(\tau)},g)$ = $e(\pi_I,g^{a(\tau)})$ $\Leftrightarrow$ $e(C/g^{r(\tau)},g)$ = $e(g^{q(\tau)},g^{a(\tau)})$ $\Leftrightarrow$ $e(g, g)^{\Psi(\tau)-r(\tau)}$ = $e(g, g)^{q(\tau)a(\tau)}$ $\Leftrightarrow$ $\Psi(X) \ =\ a(X)\cdot q(X)+r(X)$. 

\end{itemize}

\subsection{Bilinear Pairing}
We define two basic groups. $\mathbb{G}_1$, $\mathbb{G}_2$ are additive cyclic group of prime order $q$. $\mathbb{G}_T$ is a multiplicative cyclic group of prime order $q$. It is denoted as $\mathbb{G}_1 = <g>$, $\mathbb{G}_2 = <g>$ and $\mathbb{G}_T = <g>$, respectively. Thus, $e:\mathbb{G}_1 \times \mathbb{G}_2 \rightarrow \mathbb{G}_T$ is a bilinear mapping\cite{joux2000one}\cite{menezes1993reducing} with the following properties:
\begin{itemize}
    \item Bilinear: For $\forall g_1 \in \mathbb{G}_1$, $\forall g_2 \in \mathbb{G}_2$, $a, b \in \mathbb{Z}_p$. It satisfies $e(g_1^a, g_2^b)= e(g_1,g_2)^{ab}$.
    \item Non-degeneracy:$\exists g_1\in\mathbb{G}_1$, $\exists g_2\in\mathbb{G}_2$, which satisfy $e(g_1,g_2) \neq 1_{\mathbb{G}_T}$.
    \item  Compatibility: Existing an efficient algorithm for computing $e(g_1,g_2)$ for $\forall g_1 \in \mathbb{G}_1$, $\forall g_2 \in \mathbb{G}_2$.
\end{itemize}

\subsection{Mathematical Assumptions}
We elaborate the following assumptions as the basis for our scheme.
\begin{itemize}
\begin{sloppypar}
\item \textbf{n-DHE Assumption:} For one group $\mathbb{G}$ of prime order \textit{q}, let $g_i = g^{{\gamma}^i}$, $\gamma \leftarrow \mathbb{Z}_q$. On input $\left \langle g, g_1, g_2, \ldots,g_n, g_{n+2}, \ldots, g_{2n} \right \rangle \in \mathbb{G}^{2n}$, it is hard to output $g_{n+1}$.

\item \textbf{n-SDH Assumption:} Let $\alpha \in_R \mathbb{Z}_q^*$. Given as input a (n+1)-tuple $\left \langle g, g^{\alpha},\ldots, g^{{\alpha}^n} \right \rangle\in {\mathbb{G}}^{n+1}$, the probability  Pr$\left[\mathcal{A}_{n-SDH}(g, g^{\alpha},\ldots, g^{{\alpha}^n}) = \left\langle c,g^{\frac{1}{{\alpha}+c}}\right\rangle \right]$=$\epsilon(\kappa)$
 for $\forall c\in\mathbb{Z}_p\backslash \{-\alpha\}$ and every adversary ${\mathcal{A}}_{n-SDH}$.
 \end{sloppypar}
\end{itemize}

\section{ Framework and Threat Model}

Because of dynamic topology and high node mobility, IoV suffers from a range of security vulnerabilities and privacy issues. In general, vehicles should be responsible for their disseminated safety and transport messages amid adversaries or misbehaving nodes trying to forge messages for their benefits. Herein, vehicles in IoV have a two-fold requirement. Firstly, vehicles should check the legitimacy of the communicating partner; secondly, the vehicle should build their way of identification to showcase their legitimacy. Considering these requirements, we propose a privacy-preserving blockchain-based authentication protocol(PBAG). This section introduces our security framework, networks components, security assumption, and threat model.

\subsection{Framework and Components}

 We intend to provide an efficient authentication protocol in IoV. The framework of our scheme is shown in Figure 1, which comprises a three-layer architecture. The root authority in the first layer is responsible for the vehicle's certificate issuance and generating system parameters and key pairs. The second layer is composed of RTAs and blockchain, which is in charge of authenticating the vehicles or RSUs. Besides, RTAs are equipped with sufficient computing power to collect and analyze traffic messages. The third layer is composed of vehicles and RSUs. Vehicles communicate with others through the DSRC protocol. RSUs communicate with RTAs or RA via a secure transport protocol(e.g., a wired transport layer security protocol)\cite{armstrong2002dedicated}. The main components are described as follows:

{\textit{1) Root Authority(RA):}} RA should be an institution authorized by law, and it is responsible for issuing the digital certificate for vehicles. Vehicles intending to join the IoV should register themselves with the RA first. Before issuing the vehicle's digital certificate, RA thoroughly verifies the vehicle's information, including a physical examination. The information verified should include but is not limited to the driver's license, vehicle owner's identity, and the actual license plate. This procedure is usually private but could be revealed in support of legal evidence under the circumstances like accidental investigation. Besides, RA is the only entity that can track the actual ID of vehicles. In our paper, we assume the RA is a fully trusted institution and never be compromised and conspired, and it has enough computing and storage capabilities. 
 
{\textit{2) Regional Trusted Authority (RTA):}}
RTA can verify the received message from RSUs or vehicles. The blockchain is maintained by RTAs and RA. For a reasonable distribution of computing power, RTAs are responsible for collecting and analyzing traffic messages with sufficient computing power, which can help RTAs make sound responses in critical situations or optimize traffic light control. Besides, we assume the RTAs are trusted institutions, and have access to the vehicle's certificate information based on the blockchain. Therefore, RTAs are in charge of updating global commitment and certificates, which is elaborated in \ref{Certificate Update}.

{\textit{3) Road Side Units (RSUs):}}
RSUs are located alongside the roads to organize and coordinate vehicular communications in an optimized manner. After confirming identity, vehicles apply for certificates by sending a request message to nearby RSUs within their communication range. RSU is also responsible for broadcasting the latest updated global commitment to vehicles within its range and assisting vehicles to complete the anonymous proof update process. Besides, RSU also has a certificate that can prove its legitimacy. In our scheme, we assume that RSUs are semi-trusted.

{\textit{4) On-Board Unit (OBU):}}
In this paper, we assume vehicles are untrusted entities.The main computing and communication units in vehicles are OBUs. They are embedded equipment with limited computing capability, which can be used to communicate with each other within the range. We assume an OBU keeps synchronized with the RSUs. OBUs can carry out the V2V authentication without the help of RSUs. 
\begin{figure}

	\centering
	\includegraphics[width=9cm]{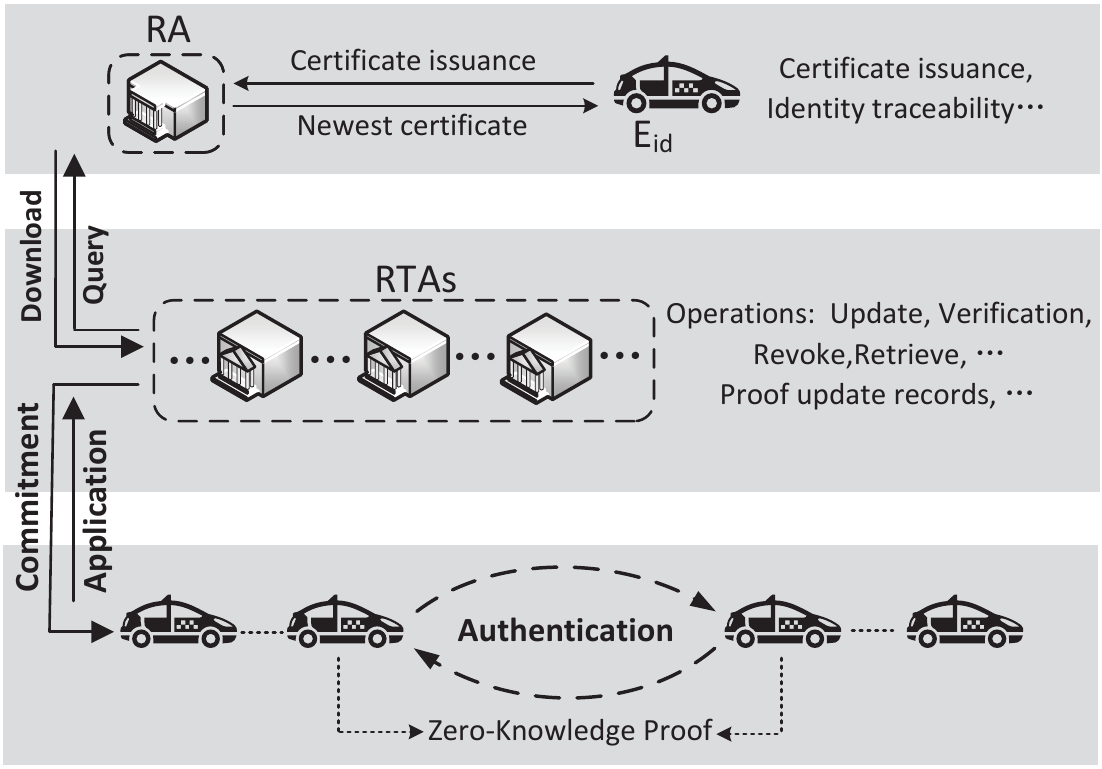}

	\label{fig2}
		{\textbf{Figure 1:} The framework of PBAG.}
		
\end{figure}

\subsection{Security Assumption and Threat Model}\label{section:Threat}
In our security hypothesis, RA and RTAs are trusted institutions, while RSUs are semi-trusted. The underlying cryptographic mechanisms used to set up a genesis block are secure for blockchain. We assume that vehicles in IoVs can be rational attackers, and they even have some prior knowledge when they launch attacks. 

The RSUs could be attacked intentionally because they are deployed in a wild area. The adversaries could collude to launch attacks like eavesdropping and stealing confidential information, which could be detected using cryptography. The malicious vehicles considered in our scheme try to obtain several certificates using the same encrypted ID to launch cyber-attacks, which can be hindered by the RA.

Under the security assumption, our model can deal with the threat arising from the authentication procedure. The common threats to our scheme are as follows:

\begin{itemize}

\item\textbf{Authentication procedure:}
The authentication protocol in IoV does not run in secure channels. The adversary could launch the attacks, particularly on the protocols. A successful attack on an authentication protocol usually does not rely on breaking the cryptographic complexity algorithm used in the protocol. Instead, the attacker obtains a certificate in an unauthorized way. We list several security issues and typical attacks below:
\begin{itemize}

\item{Correctness and integrity:} In the authentication process, the authorized vehicles are able to provide evaluation proofs that they are indeed legitimate vehicles. Besides, the management institutions can prove that messages sent by authorized vehicles are correct without being modified or forged.

\item{Traceability:} In a event of dispute, it is necessary for RA to reveal the identity by associating with the verification messages.

\end{itemize}

\item\textbf{Privacy-preserving:}  The proposed scheme must meet the following basic privacy-preserving requirements:
\begin{itemize}
    \item Anonymity: Anonymity is one of the most important aspects in wireless authentication\cite{yang2012trusted}. All vehicles can participate in the IoV communication without revealing their identity in the proposed scheme.
    \item Unlinkability: The vehicle's ability to hinder the attacker from linking the relationship between two or multiple messages that have been published.
\end{itemize}

\item\textbf{Resisting attacks:} The proposed scheme can resist the common attacks in IoV, such as the replay attack, the impersonation attack and the modification attack.
\begin{itemize}
\item{Replay attack:} An attacker repeats a previously transmitted message to intercept and retransmit its modified version, thereby fooling the honest authentication party, and gives genuine uses the illusion that they have successfully completed the authentication protocol.
\item{Modification attack:} The attacker modifies the message or authentication materials. Afterward, the attacker successfully deceives the target entity and passes its verification.

\item{Impersonation attack}. During the certificate issuance operation, adversaries are capable of forging an encrypted ID and trying to register to the RA to launch an impersonation attack where possible. Thus, the attack would hinder the certificate issuance of benign vehicles. While in the update and revocation procedure, an adversary can pretend to be the holder of the certificate and can initiate the update or revocation operation. The certificates of the benign nodes will be updated or revoked unexpectedly. In other words, the benign nodes lose control of their certificates.   
\end{itemize}
\end{itemize}

\section{The Proposed Scheme}

Vehicles transmit time-critical messages such as emergency braking and lane crossing in IoV to avoid road accidents. Thus, they should authenticate each other and then transmit the message as soon as possible to have a broader impact. To meet these requirements, we propose PBAG considering the following fundamental principles: 1) reduce latency, 2) resist attacks, and 3) preserve privacy.

 In this section, we will introduce several operations of the certificate, including issuance, update, and revocation in PBAG, which is implemented through smart contracts. Besides, we would elaborate on the authentication mechanism and a non-interactive zero-knowledge proof protocol constructed in our scheme.
 

\begin{table*}
\caption{Cryptographic Algorithms}
\label{table1}
\setlength{\tabcolsep}{4pt}
\centering

\begin{tabular}{p{120pt}|m{380pt}}
\toprule
\textbf{Notation} & \textbf{Definition}\\
\toprule
CreACC$(1^{k}) \to (npk,nsk)$ &  Generate a pair of online public key and secret key separately with a random number k \\
Encrypt($id,sk$) $\to$ $E_{id}$ & Encrypt a vehicle's $id$ with its secret key based on Elliptic Curve cryptography (ECC)\cite{johnson2001elliptic} \\
Sig$(E_{id}, sk) \to \sigma $  & A digital signature algorithm based on ECC, sign $E_{id}$ with its secret key  \\
 Check$(pk,\sigma,message) \to {0,1}$ &  Determines whether or not $\sigma$ is a valid signature on message signed by $sk$ corresponding to $pk$ \\
 $H(X)$ & Hash operation on X.\\
 $D^{\prime}(X)$ & Derivative with respect to $D(X)$.\\
Clip$(H(X)) \to \widetilde{X}$ &  A clipping function to truncate the first $n$ characters of field X \\

\toprule
\end{tabular}
\end{table*}

\subsection{Notations and Definitions}\label{Notations and definitions}
\begin{table}
\caption{Basic Notations of Certificate Operations}
\label{table2}
\setlength{\tabcolsep}{4pt}
\begin{tabular}{p{40pt}|m{195pt}}
\toprule
\textbf{Notation} & \textbf{Definition}\\

\toprule

\textbf{$E_{id}$} & Vehicle’s digital license obtained from RA. \\
\textbf{$\widetilde{E_{id}}$} & The corresponding key of the certificate stored in the database.\\
\textbf{$fpk,fsk$}& An master key pair generated by RA.\\
\textbf{$\sigma^{fsk}$}& A signature signed by an offline secret key.\\
\textbf{$T_{expired}$}& The expired time for a certificate.\\
\textbf{$t$}& Timestamp.\\
\textbf{$r$}& Random number.\\
\textbf{$P_{l}$}& Parameter set stored in the certificate.\\
\textbf{$Cer$}& Vehicle's certificate.\\
\textbf{$U_{key}$}& The update key to assist the vehicle in updating proof $\pi$.\\
\textbf{$\mathcal{E}$, $\mathcal{P}$}& The corresponding value of quantized $\widetilde{E_{id}}$,  $\widetilde{npk}$.\\

\textbf{$\omega$} & A primitive $n$-th root of unity in  $\mathbb{Z}_p$\cite{von2013modern}. \\
\textbf{$(\omega^i,u_i)$}& The parameters of vehicles for generating global commitments and performing zero-knowledge proofs, which is generated with certificate.\\
\textbf{$\pi$}& Evaluation proof utilized to verify the global commitment.\\

\textbf{$\Psi(X)$}& Global polynomial generated with total $(\omega^i, u_i)$.\\
\textbf{$C$}& Global commitment generated with global polynomial.\\

\toprule
\end{tabular}
\end{table}

In this subsection, we present several notations and definitions involved in PBAG. Above all, we choose one cryptographic hash function $H:\{0,1\}^*\rightarrow \mathbb{Z}_p$ and define a clipping function $\widetilde{X} = Clip(H(X))$ to truncate the first $n$ characters of hashed $X$ and obtain its corresponding value $\mathcal{X}$.

After that, we calculate the parameters set $P_l$ based on the certificate, including the evaluation proof $\pi$. The vehicle obtains the certificate($Cer$), including information such as $E_{id}$, $npk$, and $fpk$, after the first two stages of the system initialization process. We can utilize the clipping function to caculate $\widetilde{E_{id}}=Clip(H(E_{id}))$,  $\widetilde{npk}=Clip(H(npk))$, and corresponding value are $\mathcal{E}$ and $\mathcal{P}$. Then, we can find the unique polynomial $\Psi(X)$ for the given $n$ pairs $(x_i, y_i)_{i\in[0, n)}$ using $Lagrange$ $interpolation$ as described above. In order to more clearly exhibit the application of KZG in PBAG, we define two parameters $(\omega^i,u_i)_{i \in [0, n)}$ for each vehicle, and calculate the $\Psi(X)$, which satisfies $\Psi(\omega^i)=u_i$. Here, $\omega$ is a primitive $n$-th root of unity in $\mathbb{Z}_p$\cite{von2013modern}. Meanwhile, we define $u_i=(\mathcal{E}_i||\mathcal{P}_i)_{i \in [0, n)}$ and $u_i^{\prime}=\left[-\mathcal{RP}+ (\mathcal{E}_i||\mathcal{P}_i)\right]_{i \in [0, n)}$, where  $\mathcal{RP}$ is the value corresponding to $Clip(r\cdot npk)$ and $r$ is a random number. Based on all $(\omega^i,u_i)_{i \in [0, n)}$, the global commitment $C$ to polynomial $\Psi(X)$ can be generated in the final stage of system initialization. $\pi_i$ is a KZG commitment to $q_i(X)=\frac{\Psi(X)-u_i}{X-\omega^i}$, that is $\pi_i=g^{q_i(\tau)}$. In the authentication process, $\pi_i$ will be utilized as a proof for the vehicle to verify the global commitment $C$. Therefore, parameters set $P_l = (\omega^i,u_i,g^{\omega^i},g^{u_i},\pi_i)$ is generated and stored in the OBU.
In addition, we define the update key $U_{key}$ to assist the vehicle in updating proof $\pi_i$. Specifically, we compute $\rho_i=g^{\frac{D(\tau)}{(\tau-\omega^i)}}$, where \begin{small}$D(X)=X^n-1$\end{small} (as shown in Appendix D). And, we can obtain $\mu_i=g^{\frac{\mathcal{L}_i(\tau)-1}{\tau-\omega^i}}$, where $\mathcal{L}_i(X) = {\prod}_{j\in[0, n), j{\neq }i}\frac{X-\omega^j}{\omega^i-\omega^j}$ , and define $U_{key}=(\rho_i,\mu_i,\omega^i)$.  Finally , we assume symmetric pairings where $\mathbb{G}_1$ = $\mathbb{G}_2$ in Bilinear Pairing. The signatures involved in PBAG are generated based on elliptic curve digital signature algorithm(ECDSA)\cite{johnson2001elliptic}. The related notations and definitions are listed in Table \ref{table1} and Table \ref{table2}.


\subsection{System Initialization}
In our model, when RA has confirmed the legality of an enrolled vehicle, RA will generate the vehicle's $E_{id}$ via encrypting its real ID. Meanwhile, RA signs the vehicle's $E_{id}$ with the offline secret key and obtains $\sigma^{fsk}$, which would be utilized in certificate issuance. Vehicles accomplishing this initial process acquire the certificate from the RA. Figure 2 illustrates three phases of $E_{id}$ and certificate issuance during system initialization: Phase \uppercase\expandafter{\romannumeral1} involves RA and OBU and is performed by two parties and phase \uppercase\expandafter{\romannumeral2} and \uppercase\expandafter{\romannumeral3} need the cooperation of three parties and is accomplished in 4 steps and 2 steps, respectively.

\begin{figure}

	\centering
	\includegraphics[width=9cm]{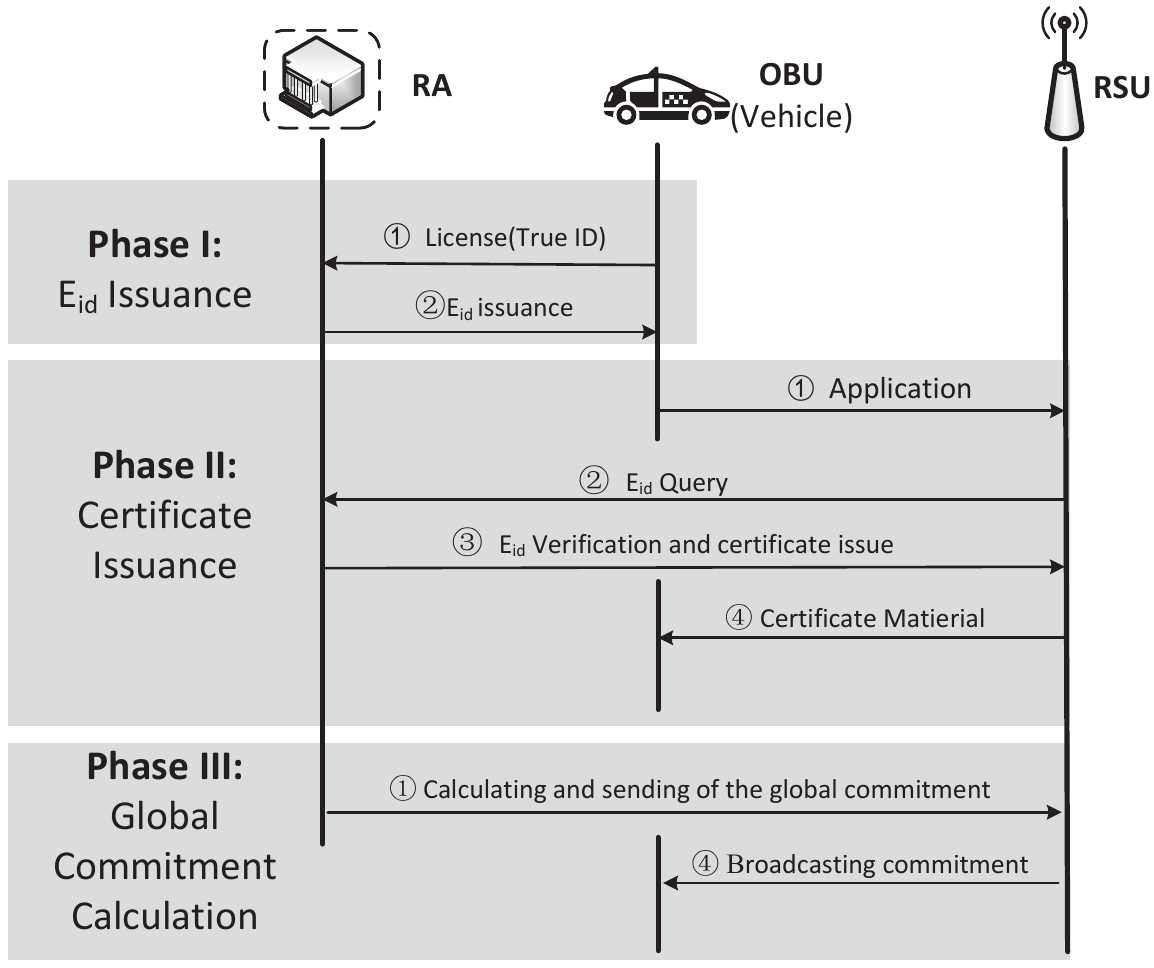}

	\label{fig2}
		{\textbf{Figure 2:} The architecture for initialization. }
		
\end{figure}

\textit{1) Phase \uppercase\expandafter{\romannumeral1}: $E_{id}$ issuance.} RA verifies the vehicle’s information and initializes the key generation procedure for each vehicle. Vehicles can obtain the encrypted digital license ($E_{id}$), offline public key ($fpk$) and signature (${\sigma}^{fsk}=sig(E_{id})$) from RA if phase I is accomplished successfully. 

master key pairs generation:
\begin{itemize}
\item RA defines an elliptic curve $\mathrm{E}$:
$y^2 = x^3 + Ax + B\ mod\ p$,
in which $p>5$ is a prime, where $A,B \in \mathbb{Z}_p$ and constants with $4A^3+27B^2 \neq 0$. Let a generator $P$ on elliptic curve $E$. Besides, the RA publishes $E$ and $P$, that is, other entities in the IoV can also obtain parameters.

\item  RA randomly chooses a random number $a\in\mathbb{Z}_p^*$ as its secret key $\textit{fsk}$, and then computes its public key $fpk= aP$, which is the master key for the RA.
\item  With the master key pairs, RA will encrypt the real ID, $E_{id}=Encrypt(id, fsk)$.
\item   RA signs the $E_{id}$ with the offline secret key,$\sigma^{fsk}=sig(E_{id}, fsk)$.  
\item   RA sends $E_{id}$, $fpk$ and $\sigma^{fsk}$ to the vehicle meanwhile reserves the offline secret key for itself.
 \end{itemize}
 
In the initialization procedure, $E_{id}$ issued for vehicles should contain enough redundancy so that other vehicles cannot reuse it. RA should retain the vehicle's confidential material and its offline secret key confidentiality. Meanwhile, a vehicle's online secret key is reserved in the OBU, which should be a tamper-proof device\cite{liu2014message}. Thus, we assume that there is no privacy disclosure and security attack risk in this phase.

\textit{2) Phase \uppercase\expandafter{\romannumeral2}: Certificate issuance.} The first operation for the vehicle is to complete the certificate issuance process in the RA. Once the $E_{id}$ is issued, the vehicle should be registered in the RA with the assistance of the nearby RSUs. The certificate issuance procedure is shown in Algorithm 1. The vehicle can obtain the certificate $Cer$ as its output with input tuple R.

 \begin{algorithm} 
\caption{: Certificate issuance procedure.}
\label{alg1}
\begin{algorithmic}[1]
\REQUIRE $R$. 
\ENSURE $Cer$.  
\STATE $R\leftarrow (E_{id},issue,npk_{t},T_{expired},\sigma^{fsk})$
\STATE \textbf{if} $Search({E_{id}})==Null$  \textbf{then}   /*check the $E_{id}$ hasn't been registered previously */
\STATE \textbf{if} $Check(fpk,\sigma^{fsk},E_{id})==1$ \textbf{then} /*validate $E_{id}$ with RA */
\STATE $P_l\leftarrow (\omega^i,u_i,g^{\omega^i},g^{u_i},\pi_i)$; /*RA calculates parameters for authentication*/
\STATE $Cer:=(E_{id},issue,npk_t, T_{expired},\sigma^{fsk})$; /*generate the certificate */
\STATE Mapping($\widetilde{E_{id}}$ $\to Cer);$ /*mapping $\widetilde{E_{id}}$ to the certificate */
\STATE \textbf{else if} {$\textit{Check}(fpk,\sigma^{fsk},E_{id})==0$} \textbf{then} 
\STATE Mapping($\widetilde{E_{id}}$ $\to pend);$ /* $E_{id}$ hasn't been validated by RA */
\STATE \textbf{else} return $\bot$;
\STATE \textbf{end if}
\STATE \textbf{end if}

\STATE \textbf{end if}
\end{algorithmic}
\end{algorithm}

\textit{Step 1-1 (Step 1 in Algorithm 1):} The vehicle sends the tuple R to the RSU located within its communication range. 
{\setlength\abovedisplayskip{0.2cm}
\setlength\belowdisplayskip{0.2cm}
\begin{equation}
 R=(E_{id},register, npk_t ,T_{expired},\sigma^{fsk})
 \end{equation}} 
 
The vehicle selects a random number $k\in \mathbb{Z}_p^*$ as its online secret key $nsk$, and calculates the corresponding public key $npk = kP$.  Therefore, we denote key generation based on ECC with a random number $k$ as: 
{\setlength\abovedisplayskip{0.2cm}
\setlength\belowdisplayskip{0.2cm}
\begin{equation}
CreACC(1^{k}) \to (npk,nsk)
 \end{equation}} 
In which the signatures $\sigma^{fsk}$  is signed by a RA’s offline secret key, which is used to prove the correctness of the vehicle’s identity. 
{\setlength\abovedisplayskip{0.2cm}
\setlength\belowdisplayskip{0.2cm}
\begin{equation}
\sigma^{fsk}=sig(E_{id},fsk) 
\end{equation} }
\textit{Step 1-2 (Step 2 in Algorithm 1):} The function Search used in step 1-2 is defined by equation (4). This function is leveraged to retrieve a vehicle’s certificate based on its encrypted identity $\widetilde{E_{id}}$. With the return value, the algorithm determines the current status of a given vehicle. Otherwise, it returns $pend$ to imply that the $E_{id}$ has not been registered.
{\setlength\abovedisplayskip{0.2cm}
\setlength\belowdisplayskip{0.2cm}
\begin{equation}
Search(\widetilde{E_{id}})=
\begin{cases}
Cer,\quad  $If the\ vehicle\ has\ certificate$ \\
revoke, \quad   $If the\ vehicle\ has\ been\ revoked$ \\
pend,  \quad     $If\ the\ vehicle\ hasn't\ been\ validated$
\end{cases}
\end{equation} }
\textit{Step1-4:}  RA calculates the parameters for authentication, which stored in parameter set $P_l$.
{\setlength\abovedisplayskip{0.2cm}
\setlength\belowdisplayskip{0.2cm}
\begin{equation}
P_l\leftarrow (\omega^i,u_i,g^{\omega^i},g^{u_i},\pi_i)
\end{equation} }
\textit{Step1-6:} After verifying the vehicle's identity, RA generates the certificate for the vehicle. In our scheme, certificates would be stored in the state database of blockchain with the smart contract, and the function $Mapping$ is constructed to achieve this procedures. Besides, the vehicle stores $Cer$ in OBU.
{\setlength\abovedisplayskip{0.1cm}
\setlength\belowdisplayskip{0.1cm}
\begin{equation}
Mapping(\widetilde{E_{id}} \to Cer)
\end{equation} }
During the issuance procedure, we have two strategies to ensure the security and efficiency of our scheme. (1) The vehicles cannot create different certificates with the same or fake identity because of unique vehicle-to-certificate mapping. Our scheme guarantees consistency for true identity and digital license, thereby ensuring the identity retention mechanism. (2) Only the RA has the offline secret key, and it is the only authority that can reveal the vehicle’s real identity.

Besides, after the issuance procedure, the vehicle selects a random number $r\in \mathbb{Z}^*_p$, and publishes $r\cdot npk$ until one authentication process is performed, that is, other entities in IoV can obtain parameter $r\cdot npk$ instead of $npk$.  After one authentication process is performed, the vehicle selects another newest random number $r^\prime$ and publishes the $npk\cdot r^\prime$. 

\textit{3) Phase \uppercase\expandafter{\romannumeral3}: Generation of global commitment.}
RA is responsible for generating the global commitment. In our scheme, we define the $\Psi(X)$ is a polynomial with coefficients $s_0, s_1,\ldots, s_d$ in $\mathbb{Z}_p$ for given $n$ pairs $(\omega^i, u_i)_{i\in[0, n)}$.  For efficiently, we assume without loss of generality that $n$ is a power of two. According to the Section \uppercase\expandafter{\romannumeral2}, RA generates a polynomial $\Psi(X)\ =\ {\sum}_{i\in[0, n)}\mathcal{L}_i(X)u_i$ using Lagrange polynomial interpolation for  vehicles under management. Therefore , we can generate a KZG commitment to $\Psi(X)$, that is $C = {\prod_{i= 0}^d}(g^{\tau^i})^{s^i} = g^{{\sum}_{i= 0}^ds_i\tau^i} = g^{\Psi(\tau)}$, which represents the state of total certificates and plays an significant role in the authentication process. Afterward, RA sends the global commitment $C$ to RSUs and RTAs, and RSU broadcasts $C$ within its management area.

\subsection{Message Authentication}
In this subsection, we demonstrate the authentication process of vehicle-to-untrusted entity as well as the authentication process of vehicle-to-trusted/semi-trusted entities. Without loss of generality, we assume participants in the authentication process are vehicles $A$(prover) and nearby vehicles, RSUs and RTAs(verifier).  
\subsubsection{The vehicle-to-untrusted entity} 
The specific authentication process of the vehicle-to-untrusted entity is as follows:
\begin{itemize}
\item \textbf{Messages generation} 
\begin{itemize}

\item [(1)] The vehicle $A$ publish the $npk_A \cdot r$ publicly, where $r$ is a random number.  The vehicle generates parameters related to message and identity as follows:
{\setlength\abovedisplayskip{0.1cm}
\setlength\belowdisplayskip{0.1cm}
\begin{align}
&E_A=nsk_A\cdot fpk\cdot r\oplus (\widetilde{E_{id}}||t)\\
&M_A= {r}_B\cdot npk_B\cdot{r} \cdot {nsk_A}\oplus H(m||t||E_A)
\end{align}}
where $r_B\cdot npk_B$ is published by the verifier. 

 \item [(2)] The vehicle $A$ calculates the related parameters of the zero-knowledge proof based on the random number $r$:
{\setlength\abovedisplayskip{0.2cm}
\setlength\belowdisplayskip{0.2cm}
\begin{align}
&\pi_{au}^A=(\pi_A)^{r-1}, g_{au}^{{\omega}^A}=(g^{{\omega}^A})^r\nonumber \\
&g_{au}^{u_A^\prime}=g^{u^{\prime}_A}\cdot({\pi_{au}^A})^{\omega^A}, g^{r-1}
\end{align}}
Therefore, this algorithm outputs $\sigma = \{\pi_{au}^A, g_{au}^{{\omega}^A},$ $g_{au}^{u_A^\prime}, g^{r-1}, E_A, M_A, m,t \}$, as the authentication tuple for the vehicle $A$ that publishes the message $m$.
  \end{itemize}


 \item \textbf{Authentication} 
  \begin{itemize}
      \item [(1)] The verifier receives the authentication tuple $\sigma$ and message $m$. It checks the freshness of the timestamp $t$ and calculates:
      {\setlength\abovedisplayskip{0.2cm}
\setlength\belowdisplayskip{0.2cm}
\begin{flalign}
&H(m||t||E_A)\oplus r_B\cdot nsk_B\cdot r\cdot npk_A\overset{?}{=}M_A & 
\end{flalign}}
If the equation holds and outputs $true$, which means valid parameter of $m$ and $(nsk,npk)$ is a valid key pair. Afterwards, the verifier would verifies that the message is sent by an authorized vehicle.
 \item [(2)] The verifier first calculates auxiliary paramete $g_{au}^{u_A} = g^{\mathcal{RP}}\cdot g_{au}^{u_A^\prime}$, where $\mathcal{RP}$ is the value corresponding to the $Clip(r\cdot npk_A)$. Therefore, the verifier can verifies:
{\setlength\abovedisplayskip{0.2cm}
\setlength\belowdisplayskip{0.2cm}
\begin{flalign}
&\quad e(C/g_{au}^{u_A}, g^{r-1}) = e(\pi^A_{au}, g^{\tau}/g^{\omega^A_{au}}) & \nonumber \\
&\Leftrightarrow e(g^{\Psi(\tau)-{u_A}-q(\tau){(r-1)}{\omega^A}}, g^{r-1}) & \nonumber \\ 
&\qquad\qquad\qquad\qquad\quad= e(g^{q(\tau)(r-1)}, g^{\tau-{\omega^A}r}) & \nonumber \\
&\Leftrightarrow e(g, g)^{\left[ \Psi(\tau)-{u_A}-q(\tau){(r-1)}{\omega^A}\right]\cdot(r-1)} & \nonumber \\
&\qquad\qquad\qquad\qquad\quad= e(g, g)^{q(\tau)({\tau-{\omega^A}r})(r-1)} & \nonumber \\
&\Leftrightarrow \Psi(\tau) - u_A = q(\tau)({\tau-{\omega^A}r}) &  \nonumber \\
&\qquad\qquad\qquad\qquad\quad+q(\tau){(r-1)}{\omega^A} &  \nonumber \\
&\Leftrightarrow \Psi(\tau) - u_A = q(\tau)(\tau - \omega^A) &  
\end{flalign}}

This algorithm can accomplish the authentication process for vehicle-to-vehicle.  If the equation holds and outputs true, which means the message is sent by an authorized vehicle.
   \end{itemize}
 \end{itemize}

 \subsubsection{The vehicle-to-trusted/semi-trusted institutions} 
The specific authentication process of the vehicle-to-trusted/semi-trusted institutions is as follows:
\begin{itemize}
\item \textbf{Messages generation} 


\begin{itemize}
\item [(1)] The vehicle $A$ generates verification parameters related to message and identity as follows:
{\setlength\abovedisplayskip{0.1cm}
\setlength\belowdisplayskip{0.1cm}
\begin{align}
&E_A=nsk_A\cdot fpk\cdot r\oplus (\widetilde{E_{id}}||t)\\
&M_A= {r}_B\cdot npk_B\cdot{r} \cdot {nsk_A}\oplus H(m||t||E_A)
\end{align}}
where ${r}_B\cdot npk_B$ is published by the verifier. 

 \item [(2)] The vehicle $A$ selects the related parameters set of the zero-knowledge proof, that is $P_l=(\omega^i,u_i,g^{\omega^i},g^{u_i},\pi_i)$. 
  Therefore, this authentication tuple $\sigma = \{P_l, E_A,$ $M_A,m,t \}$ is for the vehicle to publishes the message.
  \end{itemize}
 \item \textbf{Single authentication} 
  \begin{itemize}
      \item [(1)] The verifier receives the authentication tuple $\sigma$ and message $m$. It checks the freshness of the timestamp $t$ and verifies:
      {\setlength\abovedisplayskip{0.2cm}
\setlength\belowdisplayskip{0.2cm}
\begin{equation}
H(m||t||E_A)\oplus r_B\cdot nsk_B\cdot r\cdot npk_A\overset{?}{=}M_A  
\end{equation}}
Afterward, The verifier would check that the message is sent by an authorized vehicle.
 \item [(2)]  The verifier checks the correctness of the following equation using bilinear pairing:
{\setlength\abovedisplayskip{0.2cm}
\setlength\belowdisplayskip{0.2cm}
\begin{equation}
e(C/g^{u_i}, g) = e(\pi, g^{\tau}/g^{\omega^i})
\end{equation}}
where the relevant parameters have been introduced in the previous contents.
  \end{itemize}
 \item \textbf{Batch authentication} 
 \vspace{0.2cm}

When the verifier performs the second step of the authentication process, it can simultaneously verify messages sent by multiple authorized vehicles. Batch verification is achieved via aggregating proofs with the assistance of Partial fraction decomposition(See Appendix D). Specifically, the verifier can receive $(\omega^i,u_i,\pi_i)_{i\in I}, I\subset[0,n)$. According to previous work, given the point $(\omega^i,u_i)$, we have $\Psi(\omega^i)=u^i \Leftrightarrow \exists q_i(X), \Psi(X) - u_i = q_i(X)(X -u_i)$. The $\pi_i$ is a commitment to $q_i(X)=\frac{\Psi(X)-u_i}{X-\omega^i}$, and $\pi_I$ is the commitment to $q_I(X)$, that is $\pi_I = g^{q_I(\tau)}$. Thus, the proof aggregation is as follows:
  \begin{itemize}
      \item [(1)] The verifier calculates $R(X)$ and $D(X)$.   $R(X)$ is computed via Lagrange interpolation and satisfies $R(\omega^i)=u_i,i\in I$. $D_I(X)=\prod_{i\in I}(X-\omega^i)$.
      \item [(2)] Then, the verifier calculates $c_i=1/D^\prime(\omega^i)$ and $\pi_I=\prod_{i\in I}\pi^{c_i}_i$, which can be utilized to construct $q_I(X)=\sum_{i\in I}c_i q_i(X)$. Thus, the verifier can obtain $q_I(X) = [{\Psi(X)-R(X)}]/[{D_I(X)]}$.
      \item [(3)] Finally, the verifier can achieve the multiple evaluation proof (according to \ref{Multiple Evaluatio}):
      {\setlength\abovedisplayskip{0.2cm}
\setlength\belowdisplayskip{0.2cm}
\begin{equation}
e(C/g^{R(\tau)}, g) = e(\pi_I, g^{D_I(\tau)})
\end{equation}}
 \end{itemize}
 In short, a proof $\pi_i$ for $u_i$ is a KZG evaluation proof for $\Psi(\omega^i)$. Similarly, a sub-vector proof $\pi_j$ for $u_j$,$j\in I$, $I\subseteq[0,n)$ is a KZG batch proof for all $\Psi(\omega^i)_{i\in I}$. Thus, the verifier can verify multiple authorized vehicles simultaneously. It is worth noting that all proofs $(\pi_i)_{i\in [0,n)}$ can be computed with the assistance of the Feist-Khovratovich(FK)\cite{web2} in $O(nlog\ n)$ time. Therefore, $I$-subvector proofs can be aggregate in $O(|I|log^2|I|)$ time.

  \end{itemize}

\subsection{Certificate Update}\label{Certificate Update}
\begin{figure}
	\centering
	\includegraphics[width=9cm]{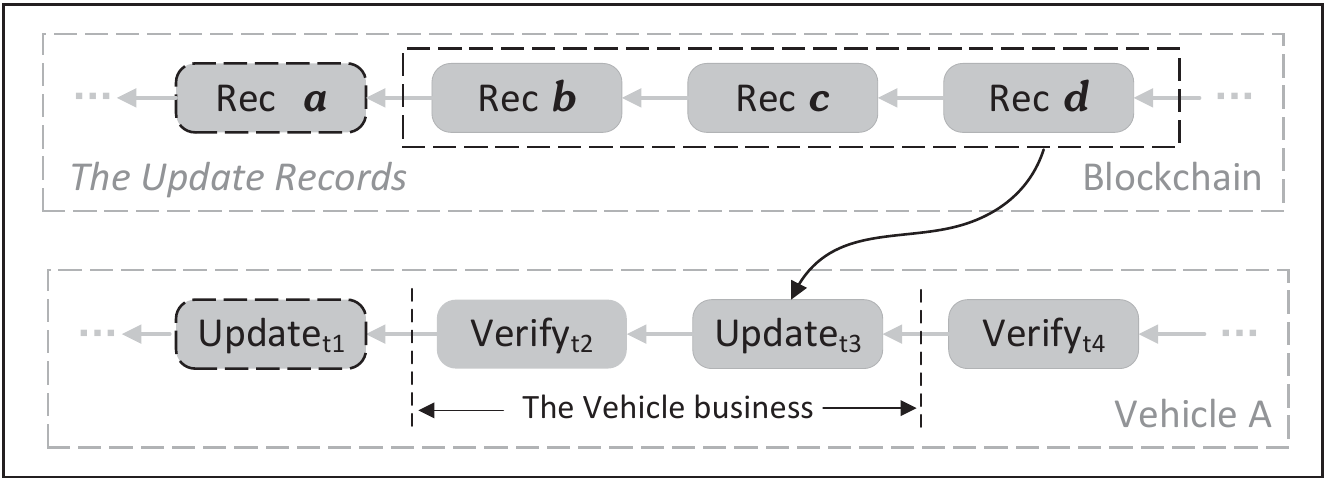}

	\label{fig2}
		{\textbf{Figure 3:} The proof update procedure.}
		
\end{figure}
In the IoV scenario, vehicles usually update their certificates under the following circumstances: (1) The identity owner initiates the update process for its own good; (2) When the vehicle intends to stay in the IoV after its certificate expired. A vehicle updates its certificate with an update strategy. Firstly, a vehicle must prove that it has been registered before. Secondly, the updated vehicle must prove that it is the holder of the previous certificate. The certificate update is divided into two phases. The Phase \uppercase\expandafter{\romannumeral1} is mainly for certificate and authentication parameters updates. The Phase \uppercase\expandafter{\romannumeral2} is mainly for the global commitment update based on the newest certificate.

\textit{Phase \uppercase\expandafter{\romannumeral1}: Certificate update.}
Before updating certificate, the vehicle must prove its ownership of the certificate, which can prevent attackers from obtaining $E_{id}$ and $npk$ through illegal means and  updating legitimate certificates arbitrarily. The vehicle first signs the new online public key with its old online secrete key, $\sigma_t^{ver}=Sig(E_{id}\|npk_{t+1},nsk_t)$, then signs the $E_{id}$  with the new secret key $\sigma_{t+1}^{nsk}=Sig(E_{id},nsk_{t+1})$, which are utilized for RTAs to verify whether the update is initiated by the owner of the previous certificate or not.  The certificate update procedure is shown in Algorithm 1.




\begin{algorithm}
 \caption{:Certificate update procedure.}
 \label{alg2}
\begin{algorithmic}[1]
\REQUIRE $ \mathcal{U}$ 
\ENSURE $Cer^{\prime}$. 
 
\STATE $ \mathcal{U} \leftarrow (E_{id},update,npk_t,npk_{t+1},T_{expired},\sigma_t^{ver},\sigma_{t+1}^{nsk})$

\STATE \textbf{if} $Search(\widetilde{E_{id}})==Cer$ \textbf{then}
\STATE \textbf{if} $Check(npk_t,\sigma_t^{ver},E_{id}||npk_{t+1}==1)$

$\&\&Check(npk_{t+1},\sigma_{t+1}^{nsk},E_{id}==1)$

$\&\&Check(npk_{t}\overset{?}{=}npk) \textbf{then}$ /* Check the certificate's ownership */
\STATE $P_l^{\prime}\leftarrow (\omega^i,u_i^{t+1},g^{\omega^i},g^{u_i^{t+1}},\pi_i^{t+1})$; /*RTA calculates newest parameters for authentication*/
\STATE $Cer^{\prime}:=(E_{id},update,npk_{t+1},T_{expired},\sigma_{t+1}^{nsk})$
\STATE Mapping ($\widetilde{E_{id}} \to {Cer^{\prime}}$); 
\STATE \textbf{if} $Search(\widetilde{E_{id}})==pend$ \textbf{then} 
\STATE $ Check(fpk,\sigma^{fsk},E_{id}){==}1$
\STATE \textbf{else} return $\bot$;
\STATE \textbf{end if} 
\STATE \textbf{end if}

\STATE \textbf{end if}
\end{algorithmic}
\end{algorithm}

\textit{Step 2-1:} The vehicle sends the tuple $\mathcal{U}$ to the RSU located within its communication range.
{\setlength\abovedisplayskip{0.2cm}
\setlength\belowdisplayskip{0.2cm}
\begin{equation}
 \mathcal{U} = (E_{id}, update,npk_{t}, npk_{t+1}, T_{expired}, \sigma_t^{ver},\sigma_{t+1}^{nsk})
\end{equation}}
The vehicle chooses another random number k, obtains its new online key pair ($npk_{t+1},nsk_{t+1}$) with  $E_{id}$.
{\setlength\abovedisplayskip{0.2cm}
\setlength\belowdisplayskip{0.2cm}
\begin{equation}
 CreACC(1^k)\to(npk_{t+1},nsk_{t+1})
 \end{equation} }
\textit{Step 2-2:} Firstly, smart contract leverages the function $Search$ to check the vehicle’s current status. Only when the return value is $Cer$, the process continues.

\textit{Step 2-3:} The RTA verifies whether the update is initiated by the owner of the previous certificate or not. RTA needs to verify two signatures $\sigma_t^{ver}$ and $\sigma_{t+1}^{nsk}$, and check whether $npk_t$ is consistent with $npk$ stored in the certificate $Cer$.
{\setlength\abovedisplayskip{0.1cm}
\setlength\belowdisplayskip{0.1cm}
\begin{align}
Check(npk_{t},\sigma_{t}^{ver},&E_{id}\|npk_{t+1})\overset{?}{=}1\\
Check(npk_{t+1},&\sigma_{t+1}^{ver},E_{id})\overset{?}{=}1\\
npk_{t}\overset{?}{=}&npk
\end{align}}
\textit{Step 2-4:} The RTA calculates newest parameters for authentication. Specifically, here is mainly the update of the three parameters, $u_i$, $g^{u_i}$ and $\pi_i$, respectively. 

For $u_i^{\prime}$, we have $\widetilde{E_{id}}=Clip(H(E_{id}))$ and $\widetilde{npk_{t+1}}=Clip(H(npk_{t+1}))$, and obtain corresponding value $\mathcal{E}_i$ and $\mathcal{P}_i^{t+1}$ and perform the following calculation:
{\setlength\abovedisplayskip{0.2cm}
\setlength\belowdisplayskip{0.2cm}
\begin{equation}
u_i^{t+1}= (\mathcal{E}_i||\mathcal{P}_i^{t+1})
\end{equation}}

Therefore, the parameter $g^{u_i}$ is also updated.

When the status of the vehicle's certificate state changes, assuming that is a change $\delta$ to $u_i$, so the newest \begin{small} $\Psi^{\prime}(X)=\Psi(X)+\delta \cdot \mathcal{L}_j(X)$ \end{small}(See Appendix C). Besides, the proof $\pi_i$ also needs to be updated. In order to express the proof update scheme explicitly, we assume that $i$ and $j$ are two vehicle codes. We will discuss both cases local proof update($i=j$) and other proof update($i\neq j$) below.

First, we discuss the case of $i= j$ and define the this update situation as local proof update. When one vehicle's certificate state changes, the proof of this vehicle would be $\pi_i\rightarrow \pi_i^{t+1}$.  We have known $\pi_i$ is a KZG commitment to $q_i(X)=\frac{\Psi(X)-u_i}{X-\omega^i}$, then:
\begin{small}
{\setlength\abovedisplayskip{0.2cm}
\setlength\belowdisplayskip{0.2cm}
\begin{equation}
    \begin{aligned}
q_i^\prime(X) &= \frac{\Psi^{\prime}(X)-(u_i+\delta)}{X-\omega^i}   \\
 & =\frac{(\Psi(X)+\delta \cdot \mathcal{L}_i(X))-u_i-\delta}{X-\omega^i}   \\
 & =\frac{\Psi(X)-u_i}{X-\omega^i}+\frac{\delta \cdot (\mathcal{L}_i(X)-1)}{X-\omega^i}   \\
 & =q_i(X)+\delta \cdot \left(\frac{\mathcal{L}_i(X)-1}{X-\omega^i}\right) 
    \end{aligned}
\end{equation}}
\end{small}
Therefore, we need a KZG commitment to \begin{small} $\frac{\mathcal{L}_i(X)-1}{X-\omega^i}$\end{small}. From \textbf{$U_{key}^i$}, we can obtain \begin{small}$\mu_i=g^\beta, \beta={\frac{\mathcal{L}_i(\tau)-1}{\tau-\omega^i}}$\end{small} and compute the updated proof: $\pi^{t+1}_i=\pi_i \cdot (\mu_i)^{\delta}$, which is uncomplicated to implement. It is worth noting that with the update of the certificate, $U_{key}^i$ and $\delta$ will be uploaded to the blockchain via a transaction.

Second, we discuss the case of $i\neq j$ and define the this update situation as other proof update. When other vehicle's certificate(e.g., vehicle $j$) state changes, the proof of vehicle $i$ would be $\pi_i\rightarrow \pi_i^{t+1}$. Figure 3 shows the update process in this case.  First, we can compute the updated quotient polynomial $q^{\prime}_i(X)$:
{\setlength\abovedisplayskip{0.2cm}
\setlength\belowdisplayskip{0.2cm}
\begin{small}
\begin{flalign}
q_i^\prime(X)
&=\frac{\Psi^{\prime}(X)-u_i}{X-\omega^i}=\frac{(\Psi(X)+\delta \cdot \mathcal{L}_j(X))-u_i}{X-\omega^i} \nonumber \\
&=\frac{\Psi(X)-u_i}{X-\omega^i}+\frac{\delta \cdot (\mathcal{L}_j(X))}{X-\omega^i} \nonumber \\
&=q_i(X)+\delta \cdot \left(\frac{\mathcal{L}_j(X)}{X-\omega^i}\right)
\end{flalign}
\end{small}}
obviously, we need a KZG commitment to $\frac{\mathcal{L}_j(X)}{X-\omega^i}$. We rewrite:
{\setlength\abovedisplayskip{0.2cm}
\setlength\belowdisplayskip{0.2cm}
\begin{small}
\begin{equation}
    \begin{aligned}
\Phi_{i,j}(X)= \frac{\mathcal{L}_j(X)}{X-\omega^i}=\frac{D(X)}{D^{\prime}(\omega^j)(X-\omega^j)(X-\omega^i)}
    \end{aligned}
\end{equation}
\end{small}} 
where \begin{small}$D(X) = \prod_{i\in[0,n)}(X-\omega^i)=X^n-1,D^{\prime}(\omega^j)=n\omega^{-j}$\end{small}. Therefore, we only need a KZG commitment  $\zeta_{i,j}$ to $\Upsilon_{i,j}(X)=\frac{D(X)}{(X-\omega^j)(X-\omega^i)}$. Based on the partial fraction decomposition, we obtain:
{\setlength\abovedisplayskip{0.2cm}
\setlength\belowdisplayskip{0.2cm}
\begin{flalign}
\qquad\quad\Upsilon_{i,j}(X)= d_1\cdot{D(X)}/(X&-\omega^j) \nonumber \\
+&d_2\cdot{D(X)}/(X-\omega^i)
\end{flalign}}
where $d_1=\frac{1}{\omega^j-\omega^i}, d_2=\frac{1}{\omega^i-\omega^j}$, obviously. Thus, we obtain $\rho_j$ and $\omega^j$ from $U_{Key}$ and compute $\zeta_{i,j}=\rho_i^{d_1}\rho_j^{d_2}$. Finally, we can compute $p_{i,j}=(\zeta_{i,j})^{\frac{1}{D^\prime(\omega^j)}}$ and update
the proof as $\pi^{t+1}_i=\pi_i\cdot(p_{i,j})^{\delta}$.

Obviously, the calculation of proof $\pi^{t+1}_i$ here follows the case of $i=j$.

\textit{Step 2-5:} The Mapping function maps the encrypted identity $\widetilde{E_{id}}$ to the new certificate $Cer^{\prime}$. Now, the vehicle stores $Cer^{\prime}$ in its local OBU.

\textit{Step 2-6:} If $Search$ returns $pend$, then RSU will inquire RTA to validate the vehicle’s identity. 

\textit{Phase \uppercase\expandafter{\romannumeral2}: Global commitment update.}
RTA is responsible for updating global commitment $C$. In the Phase \uppercase\expandafter{\romannumeral2}, the newest global $C^{\prime}$ commitment is calculated based on $C$ and newest certificate $Cer^{\prime}$. Before generating the newest global commitment, RTA needs to check validity of the proof held by the vehicle that update its certificate. Similar to previous work, $C=g^{\Psi(\tau)}$is a KZG commitment to $\Psi(X)$. For RTA, we set $l_i=g^{\mathcal{L}_i(\tau)}$\cite{2015Composable}, which can be utilized to compute $C=\prod_{i=1}^{n}(l_i)^{u_i}$ in $O(n)$ time without $\Psi(X)$. We assume that the value of $u_i$ changes by $\delta$. Therefore, the updated polynomial $\Psi(X)$ is $\Psi(X)^{\prime}=\Psi(X)+\delta\cdot\mathcal{L}_i(X)$ and the updated of commitment $C$ is $C^{\prime}=C\cdot(l_i)^{\delta}$. The newest $C^{\prime}$ is broadcast via RSUs.

\subsection{Certificate Revocation}
In the traditional authentication system, certificates expire at the time of expiry, or when it is revoked by being added to Certificate Revocation List(CRL). In our scheme, a vehicle revokes its certificate by redefining the value corresponding to the key $\widetilde{E_{id}}$ with a smart contract.

A vehicle revokes its certificate under several circumstances:1) The vehicle lost its online secret key; 2)The vehicle's certificate expired; 3) The vehicle wants to leave the network. Before a vehicle revokes its certificate, it has to be re or provide an update. A verification process should be included to prevent the vehicle from being revoked maliciously. The vehicle's certificate can be revoked by master key and online key pairs in our scheme. The revocation message $\mathcal{V}$ is initiated upon a receiving request from a given vehicle. If the vehicle loses its online secret key, it must revoke the certificate using the master key and signature. 

 \begin{algorithm}
\caption{:Certificate revocation procedure.}
\label{alg3}
\begin{algorithmic}[1]
\REQUIRE $\mathcal{V}$.
\ENSURE $\bot$. 
\STATE $\mathcal{V} \leftarrow (E_{id},revoke,npk,\sigma_t^{revoke},T_{expired})$
\STATE \textbf{if} $Search(\widetilde{E_{id}})==Cer$ \textbf{then}
\STATE \textbf{if} $Check(fpk,\sigma^{fsk},E_{id})==1$ $\&\&$

$Check(npk_t,\sigma_t^{revoke},E_{id}||T_{expired})==1$ $\&\&$

$Check(npk_t\overset{?}{=}npk)$  \textbf{then}

\STATE Mapping($\widetilde{E_{id}}$ $\to revoke)$ /*mapping the $\widetilde{E_{id}}$ to revoke */%
\STATE \textbf{else} return $\bot$; 
\STATE \textbf{end if}

\STATE \textbf{end if}
\end{algorithmic}
\end{algorithm}

\textit{Step 3-1:} The vehicle intending to revoke its certificate  sends a tuple $ \mathcal{V}$ to RSU.
{\setlength\abovedisplayskip{0.2cm}
\setlength\belowdisplayskip{0.2cm}
\begin{equation}
\mathcal{V}=(E_{id},npk_t,revoke,\sigma_t^{revoke},T_{expired})
 \end{equation}}
 The vehicle signs $E_{id}$ and $T_{expired}$ with its old online secrete key,
 {\setlength\abovedisplayskip{0.2cm}
\setlength\belowdisplayskip{0.2cm}
 \begin{equation}
\sigma_t^{revoke}=sig(E_{id}\|T_{expired},nsk_t)
 \end{equation} }
\textit{Step 3-2: }First, RA leverages the function $Search$  to check whether the vehicle’s $E_{id}$ is registered or revoked before.  
 

\textit{Step 3-3:} RSU verifies whether the revocation is initiated by the owner of the previous certificate by validating the signature $\sigma_t^{revoke}$.
{\setlength\abovedisplayskip{0.1cm}
\setlength\belowdisplayskip{0.2cm}
\begin{equation}
Check(npk_t,\sigma_t^{revoke},E_{id}\|T_{expired})\overset{?}{=}1
\end{equation}}
\textit{Step 3-4:} RA utilizes the smart contract to update the mapping $\widetilde{E_{id}}\rightarrow revoke$ to the blockchain state database.

It should be noted that after the certificate status changes, the statuses of vehicle's proof $\pi_i$ and global commitment $C$ will also change. Then the method of changing the status of $\pi_i$ and $C$ here is the same as the certificate update, just set the change $\delta$ to $u_i$. 

The certificate revocation procedure is shown in Algorithm 3. Unlike the existing schemes requiring a CRL list, our scheme leverages a smart contract to check the current status of the vehicle's certificate. Besides, the mechanism described in step 3-3 prevents the adversary from revoking certificates maliciously.

\section{Security Analysis}
We assume that the adversary in our model cannot break the standard cryptographic primitives, for instance, finding hash collisions or forging digital signatures on ECC. Further, the adversary cannot compromise the vehicles' offline and online secret keys. First, we explain how to deal with the situations when the keys are lost. Then, we demonstrate that our scheme is secure against attacks in blockchain constructions considering their operations. We also prove that our protocol proposed for the messages authentication process in IoV can resist various attacks. Lastly, we describe the level of privacy protection mechanism applied in our scheme.


\subsection{Security Proof}
\textbf{Fault-tolerance analysis: }Our proposed PBAG scheme takes advantage of the Hyperledger Fabric to realize the blockchain construction.  As we assumed in Section \ref{section:Threat}, the RTAs are trusted.  Thus, the fault-tolerant mechanism ensures the security of our scheme, and the frivolous and misbehaving registrations are hindered.  Our scheme is secure against attacks as defined in Section \ref{section:Threat}, and the details of the proof are described below.




\textbf{Security in authentication.} We define that attackers at this stage target the authentication protocol in IoV. We prove that our scheme is secure.

\textit{1) Authentication correctness and integrity:}

\textit{Proof:} The vehicle is authenticated in IoV. Vehicle generates the authentication tuple $\sigma = \{\pi_{au}^A, g_{au}^{{\omega}^A}, g_{au}^{u_A^\prime}, g^{r-1},E_A$,  $M_A,m,t \}$. Without knowing the vehicle's secret key $nsk_A$, no attacker can forge 1) a valid certificate, 2) the valid zero-knowledge proof parameters $(u_i, g^{u_i}, \pi_i)$, 3)the verification parameters related to identity and traffic messages $(E_A,M_A)$.

\textit{2) Challenges in authentication:}

\textit{Proof:} Consider there is a benign vehicle (i.e., possessing an authorized certificate) maliciously forging a certificate based on its valid key pair and trying to pass a verification equation successfully.
{\setlength\abovedisplayskip{0.2cm}
\setlength\belowdisplayskip{0cm}
\begin{equation}
H(m||t||E_A)\oplus r_B\cdot nsk_B\cdot r_A\cdot npk_A\overset{?}{=}M_A  
\end{equation}}
{\setlength\abovedisplayskip{0cm}
\setlength\belowdisplayskip{0.2cm}
\begin{equation}
 e(C/g_{au}^{u_A}, g^{r-1}) = e(\pi^A_{au}, g^{\tau}/g^{\omega^A_{au}}) 
\end{equation}}
We assume that this malicious vehicle has a forgery certificate $Cer_F$ that passes the verification equation. To pass
the verification equation, the vehicle needs to generate parameters $(u_F, g^{u_F}, \pi_F)$ to construct zero-knowledge proof corresponding to the forged certificate. However, in the authentication process, the global commitment $C$ is generated by RTA, which is based on the total authorized certificate, that is: 
{\setlength\abovedisplayskip{0.2cm}
\setlength\belowdisplayskip{0.2cm}
\begin{equation}
C = {\prod_{i= 0}^d}(g^{\tau^i})^{s_i} = g^{{\sum}_{i= 0}^ds_i\tau^i} = g^{\Psi(\tau)}
\end{equation}}
where $\Psi(X)$ is a polynomial of degree d $\leq$ n with coefficients $s_0, s_1,\ldots, s_d$ in $\mathbb{Z}_p$. $\Psi(X)$ is generated by Lagrange interpolation based on $(\omega^i,u_i)_{i\in[0,n)}$. Obviously, the computation set of global commitment $C$ does not contain $u_F$, and the vehicle cannot change the state of $C$. Thus, the vehicle cannot pass verification of  global commitment $C$ using parameters $(u_F, g^{u_F}, \pi_F)$, that is:
{\setlength\abovedisplayskip{0.2cm}
\setlength\belowdisplayskip{0.2cm}
\begin{equation}
C \neq C_F\Rightarrow e(C/g_{au}^{u_F}, g^{r-1}) \neq e(\pi^F_{au}, g^{\tau}/g^{\omega^F_{au}})
\end{equation}}

\textit{3) Traceability:}

\textit{Proof:} In our scheme, RA can track and reveal the malicious vehicles. Taking a authentication tuple and the master secret key fsk, which outputs the $E_{id}||t$ via computing the following equation. 
{\setlength\abovedisplayskip{0.2cm}
\setlength\belowdisplayskip{0.2cm}
\begin{equation}
\widetilde{E_{id}}||t = npk_A\cdot r_A\cdot fsk \oplus E_A 
\end{equation}}
The RA can compute $E_{id}$ with its secret key to identify the vehicle with the assistance of blockchain. Consequently, anyone cannot reveal the vehicle's real identity without knowing the RA's secret key.


\subsection{Privacy Analysis}
It is worthy of note that our scheme predominantly intends to preserve the privacy of the identity data. Once the vehicle's identity data is disclosed, adversaries can combine the owners' information with their vehicles. During V2V authentication, we explore every step taken to showcase the privacy-preserving property of our scheme.

\textit{Theorem 1:} {The proposed protocol is a non-interactive zero-knowledge proof on authentication tuple $\sigma$}

\textit{Proof:} First, we prove the zero-knowledge property. The authenticate tuple that the untrusted verifier receives from the prover is independent of the prover's real certificate, including parameters directly associated with the certificate, such as $E_{id}$ and $npk$.  From the certificate, the prover can obtain the parameters $(\pi,g^{\omega},g^{u^{\prime}} )$ associated with the certificate. Therefore, considering one simulator $S$: choosing random $r$, and set $\pi_{au}=(\pi)^{r-1}$, $g_{au}^{{\omega}}=(g^{{\omega}})^r$, $g_{au}^{u^\prime}=g^{u^{\prime}}\cdot({\pi_{au}})^{\omega} $. Thus, $(\pi_{au}, g_{au}^{{\omega}}, g_{au}^{u^\prime})$ is distributed correctly, and so the parameters associated with the certificate is simulated correctly. The parameters $(\pi_{au}, g_{au}^{{\omega}}, g_{au}^{u^\prime}, g^{r-1})$ in the authentication tuple $\sigma$ are related to verifying whether the vehicle is authorized in Step (2). These parameters satisfy $e(C/g_{au}^{u}, g^{r-1}) = e(\pi_{au}, g^{\tau}/g^{\omega_{au}})$, where $g_{au}^{u} = g_{au}^{u^{\prime}} \cdot g^{\mathcal{RP}}$ . For these parameters, based on the KZG polynomial commitment scheme, verifier executes the verification for global commitment $C$ using above bilinear pairing, that is, the authorization verification of prover is completed. It follows that there exists a simulator $S^{\prime}$ for this verification. The verifier obtain $\mathcal{RP}$ via preforming $Clip(r \cdot npk)$, and set $g_{au}^{u} = g_{au}^{u^{\prime}} \cdot g^{\mathcal{RP}}$. The verifier compute $e(C/g_{au}^{u}, g^{r-1}) \overset{?}{=} e(\pi_{au}, g^{\tau}/g^{\omega_{au}})$. Therefore, the bilinear pairing is simulated correctly, that is, the verification process for the authorization of the vehicle is completed.


Next, we need to prove that this protocol is a proof of knowledge. Therefore, we can define a knowledge extractor algorithm $E$, given access to a prover such that the verifier’s acceptance probability is non-negligible. If the extractor $E$ outputs the value $(\pi,g^{\omega},g^{u^{\prime}}, \pi_{au}, g_{au}^{{\omega}}, g_{au}^{u^\prime})$, such that the parameters $(\pi_{au}, g_{au}^{{\omega}}, g_{au}^{u^\prime})$ is valid. For a prover, the extractor proceeds as follows: first, it runs the extractor for the proof of knowledge protocol, and obtain the basic parameters $(\pi,g^{\omega},g^{u^{\prime}})$ from the certificate. Next, the prover can obtain the parameters $(\pi_{au}, g_{au}^{{\omega}}, g_{au}^{u^\prime})$ via $\pi_{au}=(\pi)^{r-1}$, $g_{au}^{{\omega}}=(g^{{\omega}})^r$, $g_{au}^{u^\prime}=g^{u^{\prime}}\cdot({\pi_{au}})^{\omega} $. Then:
{\setlength\abovedisplayskip{0.2cm}
\setlength\belowdisplayskip{0.2cm}
\begin{equation}
    e(C/g_{au}^{u}, g^{r-1}) =e(\pi_{au}, g^{\tau}/g^{\omega_{au}})
\end{equation}}
And based on $g_{au}^{u} = g_{au}^{u^{\prime}} \cdot g^{\mathcal{RP}}$, therefore the tuple $\widetilde{P_l}=(\pi_{au}, g_{au}^{{\omega}}, g_{au}^{u^\prime})$ satisfies the above equation and hence are the blind parameters on $(\pi,g^{\omega},g^{u^{\prime}})$, so our extractor outputs $(\pi,g^{\omega},g^{u^{\prime}}, \pi_{au}, g_{au}^{{\omega}}, g_{au}^{u^\prime})$.

Finally, our protocol is a non-interactive authentication scheme. There are no interaction between the prover and verifier during the authentication process, which is suitable for the wireless scenario.

\textit{1) Anonymous:}

\textit{Proof:} The property of anonymity is achieved in three ways, 1) The combined value generated based on $Cer$, $E_{id}$, $npk$ is used in the authentication process instead of the certificate or $E_{id}$, 2)  the membership
public key $npk$ is randomized by random number $r$, 3)the authentication is implemented with non-interactive zero-knowledge proof. We can give a proof sketch for anonymity. The adversary cannot recover certificate or $E_{id}$ from the hash and truncated value. Besides, the random number $r$ and public key($npk$) are non-public, and the adversary cannot parse the public key and random number from the randomized public key. Based on the \textit{n-DHE} assumption and \textit{n-SDH} assumption, the adversary cannot compute the random number or the information to be able to tag the vehicle(e.g., $\omega^i, u_i$) from the authentication tuple.

\textit{2) Unlinkability:}

\textit{Proof:} The proposed scheme provides unlinkability that no verifier can tell whether two authentication tuples were derived from the same vehicle. For the  authentication tuple $\sigma = \{\pi_{au}^A, g_{au}^{{\omega}^A}, g_{au}^{u_A^\prime}, g^{r-1},E_A, M_A,m,t \}$, each element is generated with random number $r$. Obviously, the attacker can derive evaluation proof $\pi$ from the same vehicle if it obtains the authentication tuple initiator's secret key $nsk$ or breaks the \textit{n-DHE} assumption.

\subsection{Resistance to Attacks}
We analyze several common attacks in IoV and illustrate that our scheme can resist these attacks.

\textit{1) Impersonate attack}.

\textit{Proof:} According to the registration algorithm, the certification posted to the blockchain is composed of a tuple $\mathcal{R}$=$(E_{id}, register, npk,T_{expired},\sigma^{fsk})$. If an adversary $\mathcal{A}$ tries to forge a certificate, there are several routines to follow. First, $\mathcal{A}$ attempts to figure out the real ID  of a benign vehicle. The real ID is issued and reserved by the RA after a thorough examination in our scheme. There is no way for $\mathcal{A}$ to access the real ID. Second, $\mathcal{A}$ aims at $E_{id}$, which is issued by RA but reserved in the vehicle’s OBU. According to \cite{intelligent2013ieee}, a tamper-proof device installed in the OBU is responsible for storing confidential material. Third, the RSUs usually request the RA to perform a validity check during the registration. The fake ID and $\widetilde{E_{id}}$ cannot pass this validity check. Thus, we can conclude that our scheme can resist impersonate attacks during registration.

\textit{2) Replay attack:}

\textit{Proof:} An adversary $\mathcal{A}$ launches an attack by replaying the critical identity information between the participants involved in the authentication process. In our scheme, the global polynomial commitment $C = g^{\Psi(X)}$ is dynamic following each vehicle certificate's state. Therefore, for realizing the correspondence between the proof and the global certificate status, the proof (e.g., $\pi^{{A}})$ should be updated before the vehicle implements the authentication process, as shown in \ref{Certificate Update}. It is worth noting that information required for the proof update is hidden(e.g., $\omega^A$). Therefore, an attacker cannot utilize the old proof to pass the verification based on the latest commitment and cannot update the old proof. On the other hand, the tuple $\sigma = \{\pi_{au}^A, g_{au}^{{\omega}^A}, g_{au}^{u_A^\prime}, g^{r-1},E_A, M_A,m,t \}$ sent by authentication launcher includes timestamp $t$, therefore, parameter $\sigma$ changes with $t$, which prevents the outdated message and resist the replay attack.

\textit{3) Modification attack:}

\textit{Proof:} According to Theorem 2, the modification of the authentication tuple $\sigma = \{\pi_{au}^A, g_{au}^{{\omega}^A}, g_{au}^{u_A^\prime}, g^{r-1},E_A, M_A,m,t \}$ cannot pass the verification equation(31)(32). Therefore, our scheme can resist the modification attack.


\section{Performance Evaluation}\label{Performance Evaluation}
In this section, we evaluate the performance of our scheme through extensive simulations, including  analyzing the authentication overhead and evaluating the practical viability of our proposed scheme against the existing state-of-the-art approaches.

\subsection{Experiment Settings}

We simulate PBAG on a machine with an AMD Ryzen 7 5800H with Radeon Graphics CPU @ 3.20GHz CPU and 16GB RAM. The prototype is implemented in Hyperledger Fabric v2.0.0 with Raft consensus, and its chaincodes  are developed in Golang\cite{golang}. 

Furthermore, we have two organizations, which are RA and RTAs. RA and RTAs are utilized as endorsing peers. Following the endorsement policy, when more than $2n+1 $signatures from the endorsing peers are valid, one transaction will be successfully committed. It is worth noting that the vehicle in our scheme would be not as the endorsing peer and have no rights to access the channels, due to its mobility and limited computing power. In the simulation, we construct four peer nodes belonging to two organizations, namely $org1$ and $org2$, three order nodes, and one client node in one channel.

In our scheme, we implement bilinear pair programs with the MIRACL library. We construct the symmetric pairing $e: \mathbb{G}\times\mathbb{G}\rightarrow\mathbb{G}_T$, which is built on the security level of 80 bits. $\mathbb{G}$ is an additive group generated by a point $\hat{p}$ with the order $\hat{q}$, where $\hat{p}$ is the prime number $\hat{p}$ = 512 bits and $\hat{q}$ is a 160 bits prime number. ECC is constructed at 128 bits security level. In order to simulate the average communication latency between RTAs and vehicles, the ns-2.35 is used with the communication protocol IEEE 802.11p and the routing protocol AODV. 


In order to estimate time consumption of certificate operations, we run the smart contracts with various number of vehicles in the network scaled at$10$, $10^2$, $10^3$, and $10^4$. In the experiment, we combine the vehicle's plate with VIN to form an 18-digit number as their real $ID$ parameter. With this $ID$, we apply the crypto series library in Golang to generate $E_{id}$ and master key pairs.

\subsection{Overhead of blockchain construction}
We evaluate the overhead of blockchain construction by measuring throughput of transactions in the prototype. Throughput is defined as the rate that transactions are committed to the ledger\cite{thakkar2018performance}. We calculate the average measured as throughput during the steady state of our simulations. For example, vehicle revocation is implemented by calling the smart contract with a new transaction, which aims to appending the revocation data to the blockchain state database. Certificate issuance and update are similar. Therefore, we pre-determined the transaction size to be 4kb, and the block size is set to contain 10, 25 and 40 transactions respectively. In the simulation, we set the RTA to receive transactions at a rate of 20 to 140 per second for write operation.  The simulation results are depicted in Figure 4. For the block with different sizes, we can observe that throughput improves in a near-linear trend with the transaction arrival rate increase. 

\begin{figure}
	\centering
	\includegraphics[width=8.5cm]{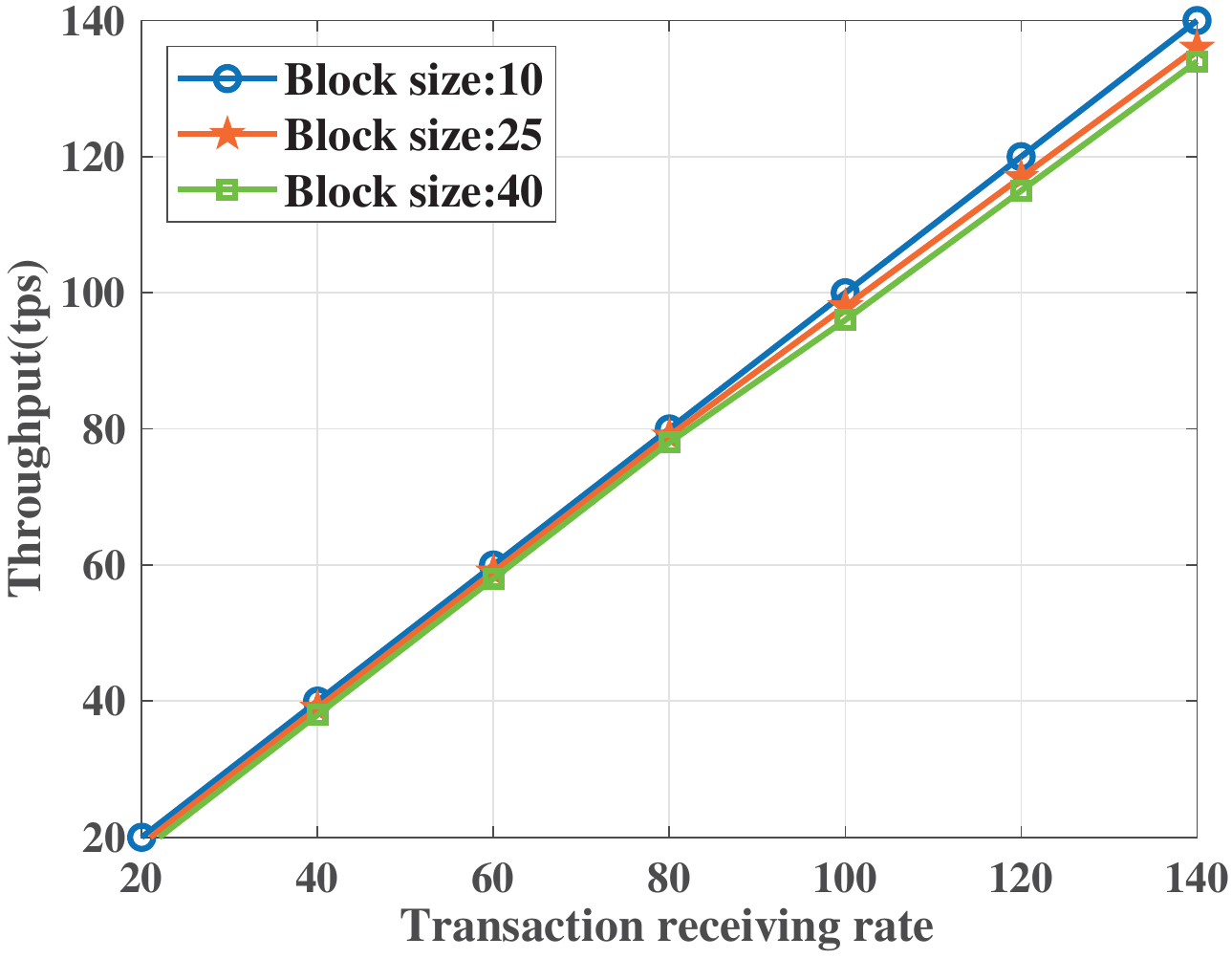}
	\label{fig4}
	
		{\textbf{Figure 4:} Transaction throughput for certificate operation in Hyperledger Fabric.}
\end{figure}

\subsection{Certificate Operation}
\begin{figure*}
\begin{center}
\subfigure[]{
\includegraphics[width=8.3cm]{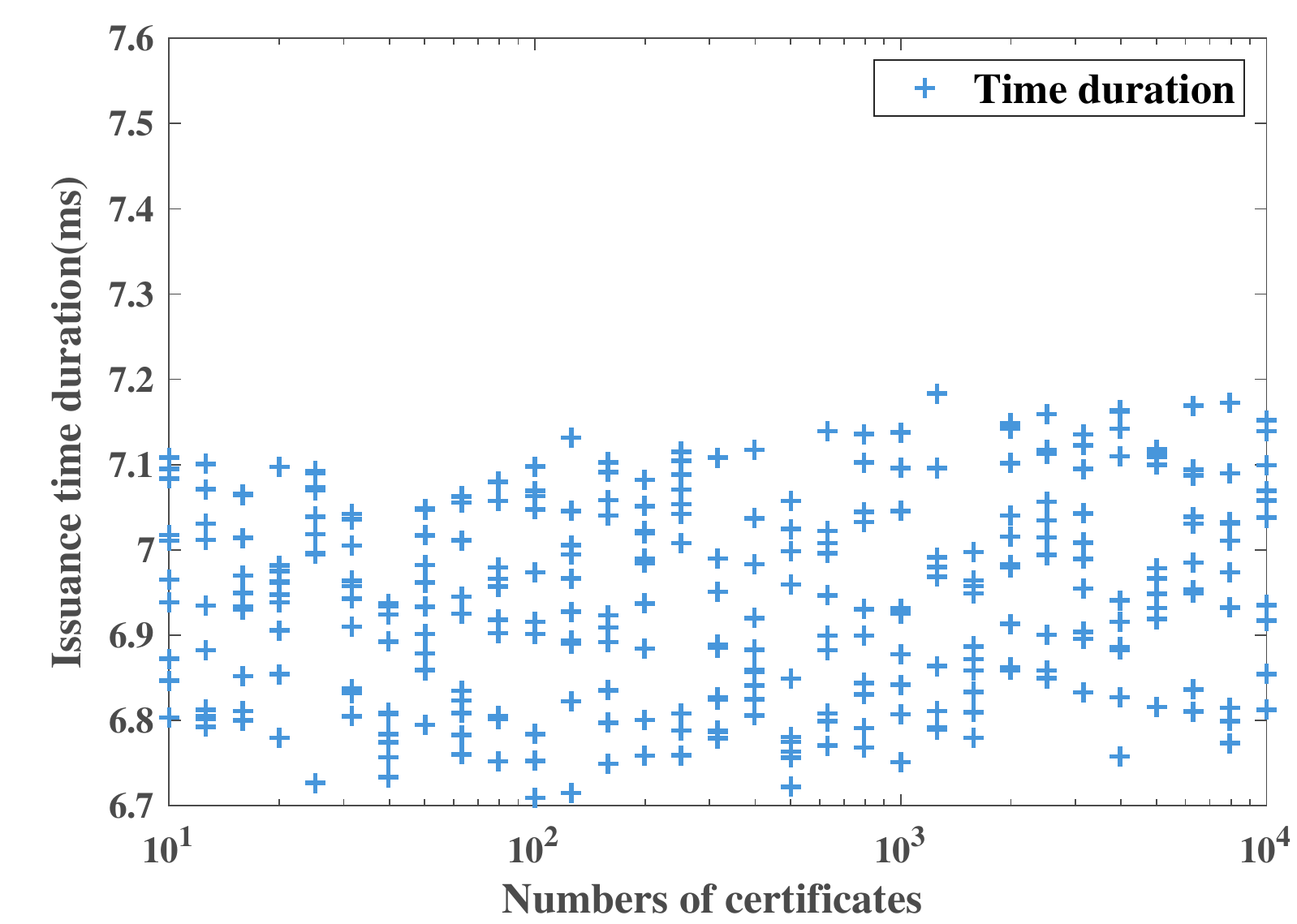}
}
\subfigure[]{
\includegraphics[width=8.3cm]{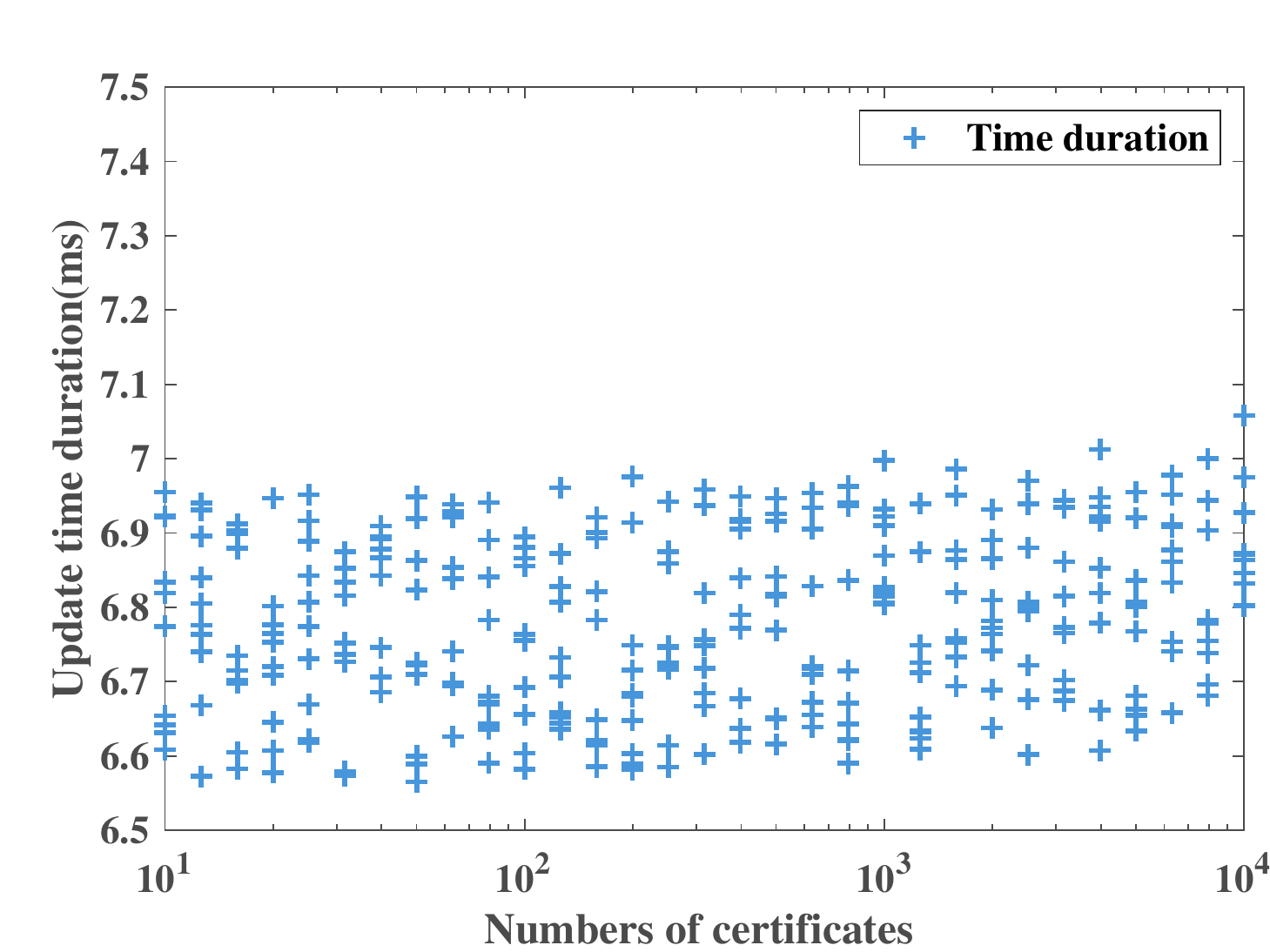}
}
\subfigure[]{
\includegraphics[width=8.3cm]{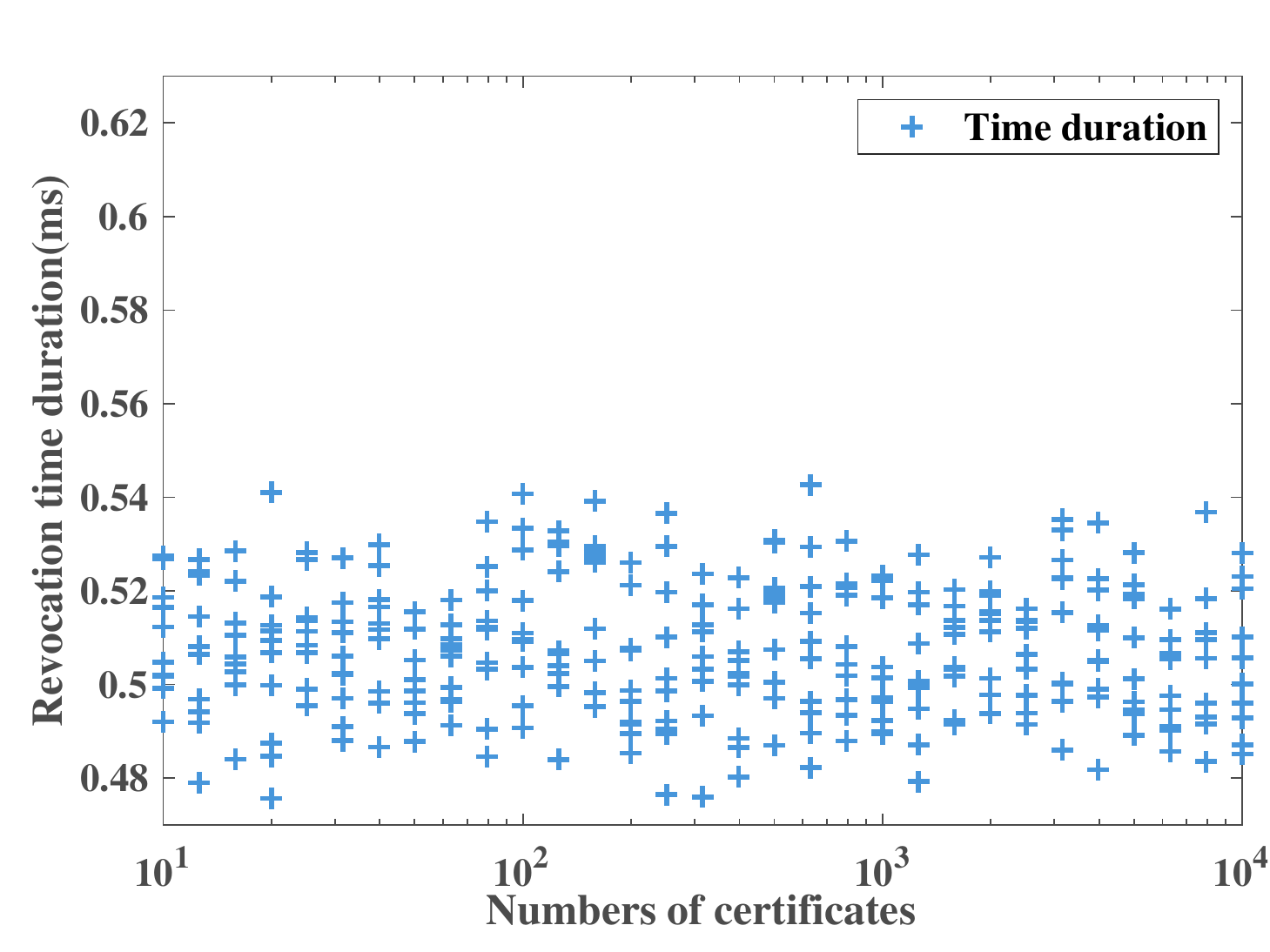}
}
\subfigure[]{
\includegraphics[width=8.3cm]{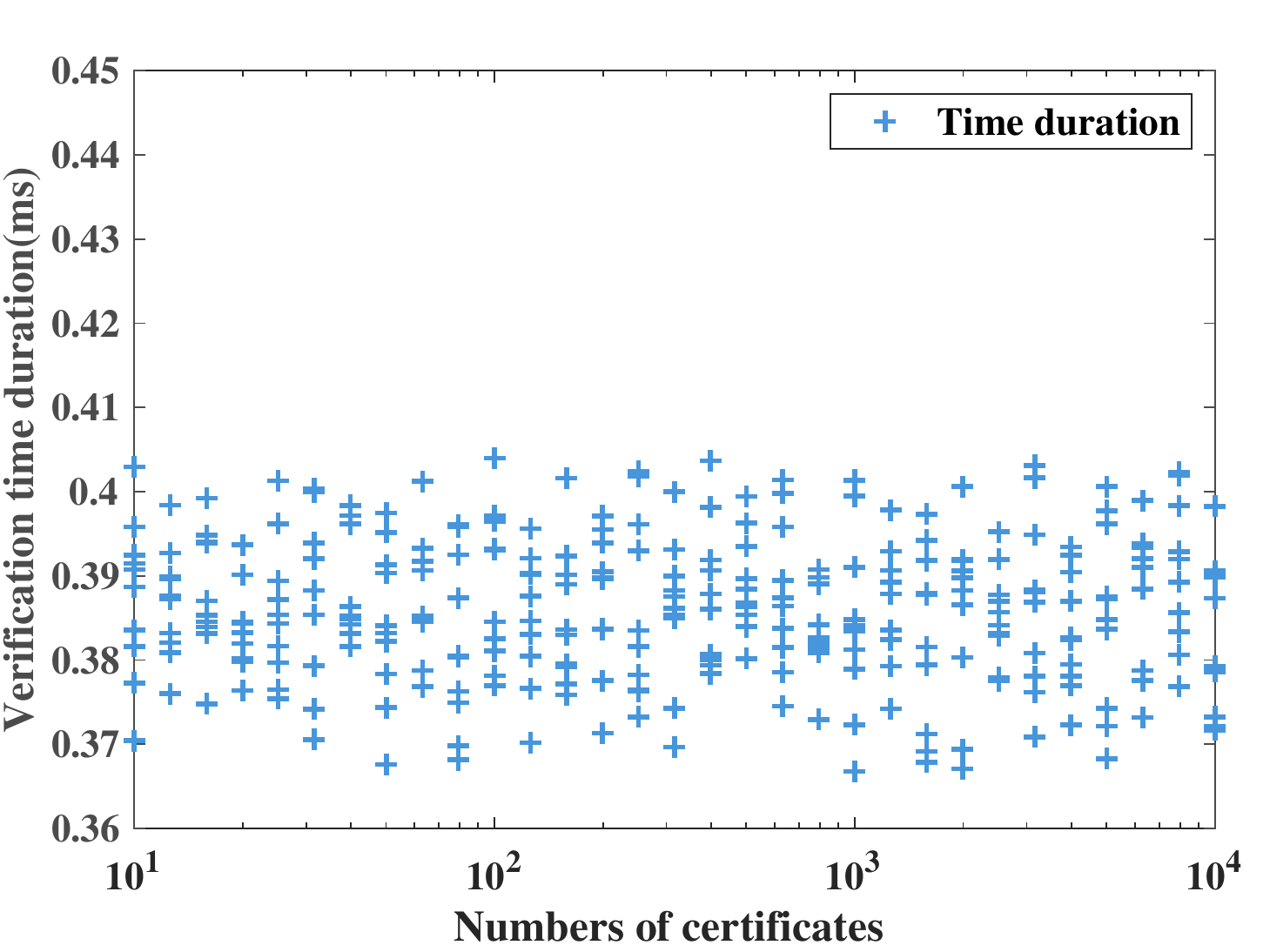}
}
\end{center}
\begin{center}
\textbf{Figure 5.} The time duration of certificate operations. The fig.(a), fig.(b),fig.(c) present the time duration of certificate issuance, update, revocation, respectively. Fig. (d) presents the time overhead of verification involved in the above operations. The number of vehicles in the simulation scaled at $10$, $10^2$, $10^3$, and $10^4$.
\end{center}
\end{figure*}


We carry out simulations to estimate the time consumption of the certificate operations in PBAG with smart contracts, as presented in Figure 5. Specifically, Figure 5 (a) and (b) show the time incurred for calling the issuance and updating contracts, respectively. Figure 5 (c) shows the time duration for calling revocation with different certificate scales. All certificate operations involve validating whether the vehicle has a legal signature and an online key. Therefore, we simulate this verification process additionally, and the related results are shown in Figure 5 (d). We observe that the data presented in each subfigure did not alter significantly, and the data is distributed in a specific interval. According to our simulation, the average time consumption for the contract issuance is 6.95ms, while verification is 0.39ms. This experimental result is due to the issuance and update process needed to regenerate the certificate, including the key generation process, which is relatively time-consuming. Figure 5 indicates that by withstanding small  fluctuations when between by running the smart contract multiple times in the Fabric, and our scheme is technically feasible according to the cost of time. 



\begin{figure*}[htb]
\begin{center}
\subfigure[]{
\includegraphics[width=8.3cm]{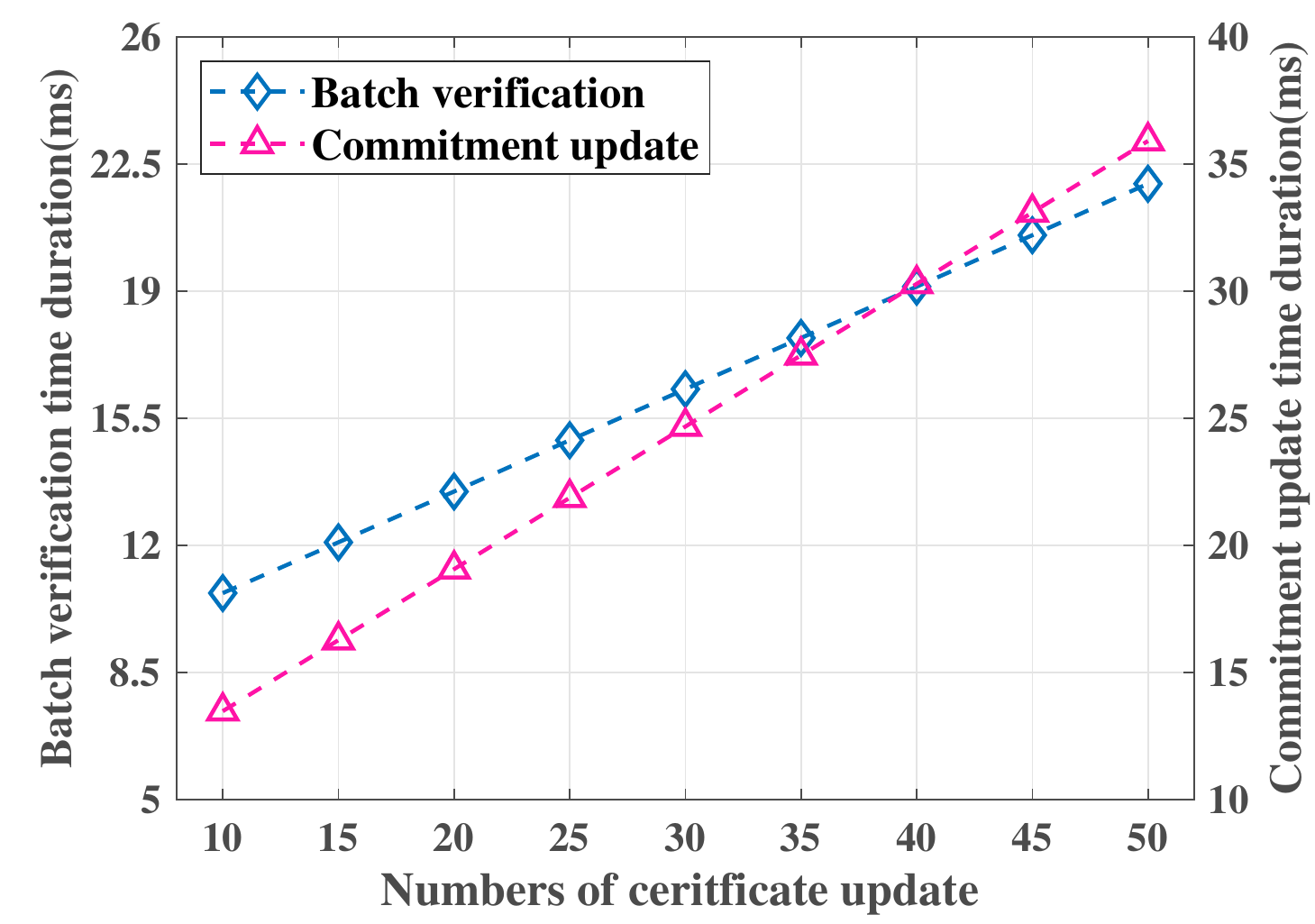}
}
\subfigure[]{
\includegraphics[width=8.3cm]{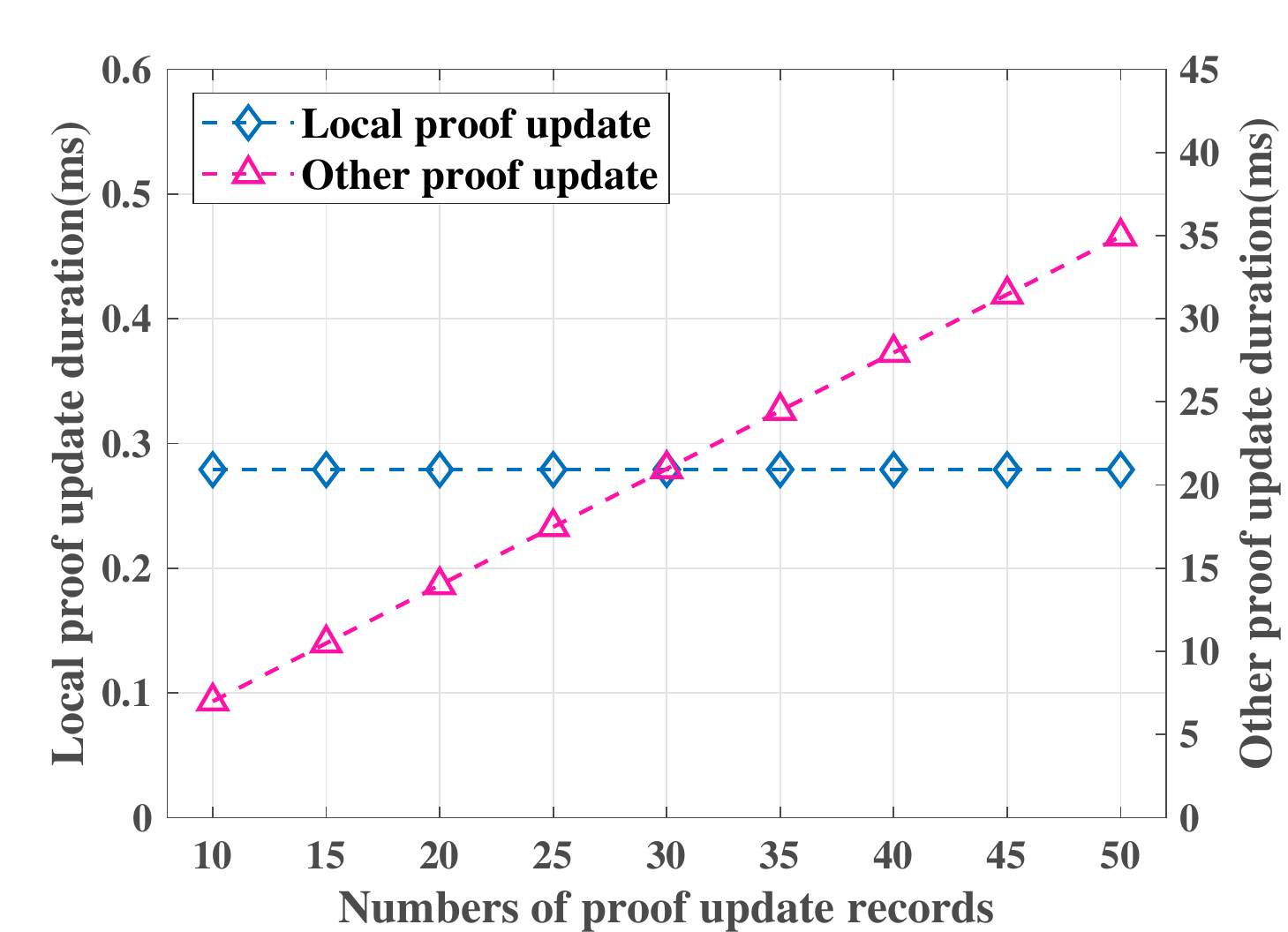}
}
\end{center}
\begin{center}
\textbf{Figure 6.} The time overhead of commitment and proof update. Fig.(e) presents the time duration of global commitment update, and the batch verification operation time involved in the global commitment update process. Fig.(f) presents the time duration of local proof update($i=j$) and other proof update($i \neq j$).
\end{center}
\end{figure*}

Besides, the authentication is accomplished based on zero-knowledge proof in our scheme. There are three crucial elements involved in the zero-knowledge proof protocol: certificate, proof, and global commitment. Proof and global commitment are elements utilized to implement authentication by the verifier through bilinear pairing, and both need to be updated following the state of the certificate(elaborated in \ref{Certificate Update}).

\begin{itemize}


\item \textbf{Update of global commitment} 
The update of global commitment is divided into two phases, the verification of evaluation proof and the calculation of the newest global commitment. We simulate the update process with smart contracts as shown in Figure 6(e). We can observe that time duration curves of both phases have an upward trend as numbers of certificate update increasing. Based on the batch-enabled mechanism, when there are 50 certificate update records, the time overhead of verification is 21.956ms, and the calculation of the newest global commitment can be accomplished within 35.906ms. 


\item \textbf{Update of evaluation proof}
For vehicles, the situation of proof update can be divided into two categories, local proof update and other proof update. We implement the simulation with smart contracts to estimate the update performance, as shown in Figure 6(f). For local proof update, we can observe that the time overhead is basically constant at 0.279ms. Besides, the time duration curve shows an upward trend for other proof updates as the number of proof update records increases. The time duration is about 34.95ms when there are 50 proof update records.


\end{itemize}


\subsection{Computation Cost Analysis and Comparison}

\begin{sloppypar}
Here, we analyze the computation cost of our proposed PBAG scheme, against the state-of-the-art schemes, including EADA\cite{yang2020delegating}, RCoM\cite{wang2018privacy}, P2BA\cite{feng2021p2ba}. First, we calculate the execution time of some basic cryptographic operations by using the MIRACL library. For the accuracy of the evaluation, we run the cryptographic operations 1000 times and take the average values as the final results.  The execution time of the necessary cryptographic parameters and the authentication overhead of some related schemes are shown below:
\end{sloppypar}

\begin{itemize}
\item{\textbf{$T_{bp}$}} is the time required to perform a bilinear pairing, $T_{bp}\approx 3.803$ms
\item{\textbf{$T_h$}} is the  time required to implement the hash function, $T_h\approx0.001$ms
\item{\textbf{$T_{sm}$}} is the  time required to implement one scalar multiplication under $\mathbb{G}_1$, $T_{sm}\approx0.141$ms.
\item{\textbf{$T_{ex}$}} is  time required to implement one exponentiation in $G_T$, $T_{ex}\approx0.138$ms.
\item{\textbf{$T_{mpt}$}} is  time required to implement one MapToPoint hash operation of the bilinear pairing, $T_{mpt}\approx0.092$ms.
\end{itemize}

\begin{table*}
\caption{Comparison For Computational Costs}
\label{table4}
    \begin{threeparttable}
\setlength{\tabcolsep}{4pt}
\centering
\begin{tabular}{p{60pt}|m{120pt} | m{160pt} | m{160pt}}
\toprule
\textbf{Scheme} & \textbf{Generate one message} & \textbf{For one message authentication }& \textbf{For $n$ messages authentication }\\
\toprule
RCoM &  $7T_{bp}+4T_{ex}$ & $(2s+6)T_{bp}+7T_{ex}$ & $(2s+6)nT_{bp}+7nT_{ex}$ \\

EADA & $2T_{bp}+3T_{sm}+3T_{mtp}$ & $4T_{bp}+3T_{sm}+4T_{mtp}$ & $(3n+1+\kappa)T_{bp}+(n+2)T_{sm}+(3+n+\kappa)T_{mtp}$ \\

P2BA  & $2T_{bp}+11T_{sm}+12T_{ex}$ & $4T_{bp}+10T_{sm}+10T_{ex}$  & $4T_{bp}+(6n-1)T_{sm}$\\
Our proposal & $3T_{ex}+T_{h}$ & $2T_{bp}+2T_{sm}+T_{h}$ & $2T_{bp}+nT_{sm}+(n+2)T_{ex}+T_L+nT_h$\\
\toprule
\end{tabular}

\begin{tablenotes}   
        \footnotesize      
        \item[1] $T_L$:The time to calculate using Lagrangian interpolation.
        
        \item[2] $\kappa$:number of compromised edge nodes in EADA.

        \item[3] s:number of equivalence class in RCoM.

      \end{tablenotes}         
    \end{threeparttable}
    
\end{table*}

For describing the authentication process accurately, we calculate the computation cost of the various phases in the authentication operation. First, we calculate the computation cost for generating one authentication message. For one vehicle, it mainly needs to perform 3 exponentiation operations and 1 hash operation. Thus, the message generating time per vehicle is 0.415 ms. Afterward, we evaluate the performance of the RTA/RSU's message verification phase. The infrastructure needs to perform 2 bilinear pairing, 2 scalar multiplication, and 1 hash operation when verifying a single message. Thus, the authentication time per vehicle is 7.889ms. Finally, our scheme supports batch message verification, which significantly facilitate the authentication performance. We can compute that verifying 100 messages synchronously needs about 36.171 ms. Therefore, our scheme satisfies the requirement latency to support applications in IoV.


Table \uppercase\expandafter{\romannumeral3} gives the comparison of computation overhead among  recently proposed schemes approaches including EADA, RCoM, P2BA. The primary operations involved in the related schemes are $T_{bp}$, $T_{sm}$, $T_{ex}$ and $T_{mtp}$. For authentication with the EADA scheme, the cryptographic operations require to perform 4 bilinear pairing operations, 3 scalar multiplication under $\mathbb{G}_1$, 4 MapToPoint hash operations of the bilinear pairing, respectively. Thus, the total computation cost is $4T_{bp}+3T_{sm}+4T_{mtp} \approx 16.003$ ms. For authentication with the RCoM scheme, the cryptographic operations require performing $2s+6$ bilinear pairing operations, 7 exponentiation under $\mathbb{G}_T$, respectively. Thus,  the total computation cost is $(2s+6)T_{bp}+7T_{ex} \approx 38.996$ ms with number of equivalence class assumed to be 5. For authentication with 
P2BA scheme, the cryptographic operations require to perform 4 bilinear pairing operations, 10 scalar multiplication under $\mathbb{G}_1$,  10 exponentiation under $\mathbb{G}_T$ of the bilinear pairing, respectively. Thus, the total computation cost is $4T_{bp}+10T_{sm}+10T_{ex} \approx 18.002$ ms. Therefore, our scheme outperforms above at least 50.70$\%$, 79.77$\%$, 56.18$\%$, respectively, for EADA, RCoM, and P2BA.

In order to highlight the efficiency of the proposed PBAG, we compare the computation times of $n$ authentication in the proposed scheme with three state-of-art schemes, EADA, RCoM, P2BA, as shown in Figure 7. It should be noting that $T_L$ is the time to calculate using Lagrangian interpolation, which is 0.0179ms, 0.0811ms, 0.1696ms, 0.2237ms, 0.2893ms, 0.3672ms, respectively for the 20, 40, 60, 80, 100, 120 vehicles.  Specifically, taking 100 authentication requirements as an example, the authentication delay is 1180.25ms, 3899.6ms, and 99.671ms respectively for EADA($\kappa$=3), RCoM(s=2), P2BA. Compared with the other three schemes, the time delay of the PBAG scheme is 36.171ms, which outperforms above at least 96.94$\%$, 99.07$\%$, 63.7$\%$, respectively for EADA, RCoM and P2BA. Therefore, the proposed scheme is more efficient than the comparison schemes.


\begin{figure}
	\centering
	\includegraphics[width=9cm]{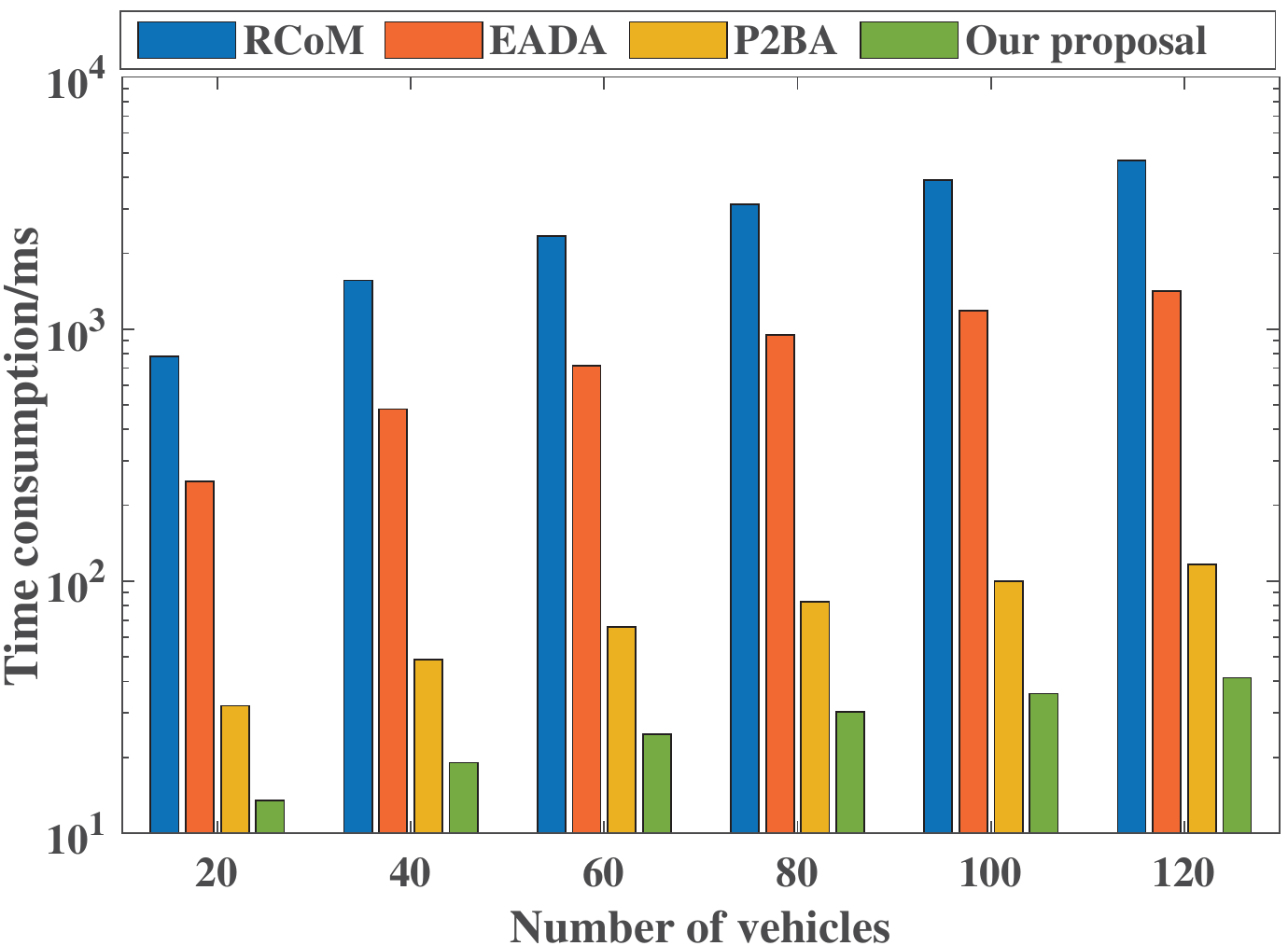}
	\label{fig2}
	
		{\textbf{Figure 7:} Comparison of computation cost.}
\end{figure}

\subsection{Authentication Latency and Comparison}

\begin{table*}[htbp]
  \centering
    \begin{threeparttable}

  \caption{Comparison of Authentication Message Size}
    \begin{tabular}{p{60pt} p{70pt} p{70pt} p{90pt} p{60pt} }
    \toprule
    \multicolumn{1}{c}{\multirow{2}[0]{*}{Scheme}} & \multicolumn{2}{c}{{Verify a message}} & \multicolumn{2}{c}{{Verify $n$ message}} \\
         & \multicolumn{1}{c}{vehicle $\rightarrow$ verifier} & \multicolumn{1}{c}{verifier $\rightarrow$ vehicle} & \multicolumn{1}{c}{vehicle $\rightarrow$ verifier} & \multicolumn{1}{c}{verifier $\rightarrow$ vehicle} \\
    \midrule
\makecell[c]{RCoM}    &    \makecell[c]{1176 bytes}  &     \makecell[c]{N/A} &\makecell[c]{1176n bytes}&   \makecell[c]{N/A}  \\
\makecell[c]{EADA}    &    \makecell[c]{452+964$\kappa$ bytes}  &     \makecell[c]{256 bytes} &\makecell[c]{452+964$\kappa$+580n bytes}&   \makecell[c]{256n}  \\
\makecell[c]{P2BA}    &    \makecell[c]{768 bytes}  &     \makecell[c]{N/A} &\makecell[c]{768n bytes}&   \makecell[c]{N/A}  \\
\makecell[c]{Our proposal}    &    \makecell[c]{554 bytes}  &     \makecell[c]{N/A} &\makecell[c]{554n bytes}&   \makecell[c]{N/A}  \\
    \bottomrule
    \end{tabular}%
    
    \begin{tablenotes}   
        \footnotesize      
        \item[1] $\kappa$:number of edge nodes in EADA.

      \end{tablenotes}         
    \end{threeparttable}

  \label{tab:addlabel}%
\end{table*}%
In this subsection, we analyze the authentication delay based on the more complex authentication scenario that vehicle to RTA/RSUs. We define authentication latency containing three main parts, the computation time of the vehicle $T_{gen}$, the message propagation time $T_{p}$, and the computation time of the verifier $T_{ver}$. Therefore, the total authentication latency is $T_{gen}+T_{p}+T_{ver}$ involved in the process from the vehicle generating a message to when the message is successfully verified. Among them, $T_{gen}$ and $T_{ver}$ have been elaborated in the previous subsection. We will analyze the calculation of $T_{p}$ detailedly in the following contents. Besides, for accurately calculating the $T_{p}$, we perform the simulation based on the road scenario in ns-2.35.

There are three influencing factors in the message propagation process, message size, vehicle speed, and vehicle density. First, we analyze the message size in our proposal and related schemes. The size of message $\sigma = \{\omega, u,g^{\omega},g^{u},$ $\pi, E_A, M_A,m,t \}$ sent by vehicles is 554 bytes. Besides, we compare message size at various stages of authentication with related schemes, which is presented in the Table \uppercase\expandafter{\romannumeral3}. We can observe that our proposal outperforms other schemes.

\begin{figure}
	\centering
	\includegraphics[width=9cm]{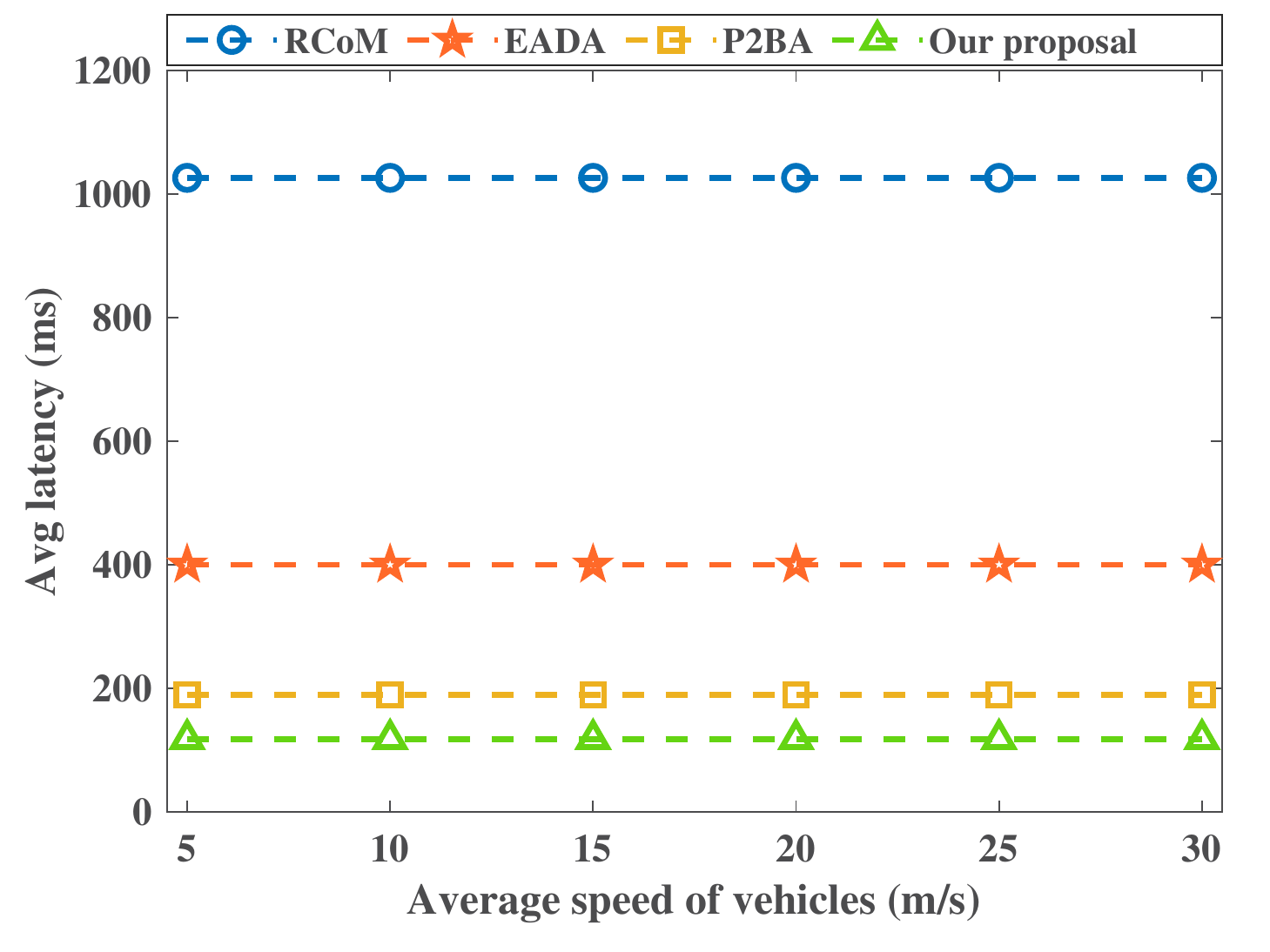}
	\label{fig2}
	
		{\textbf{Figure 8:} Comparison of authentication latency at different vehicle speeds.}
\end{figure}

\begin{figure}
	\centering
	\includegraphics[width=9cm]{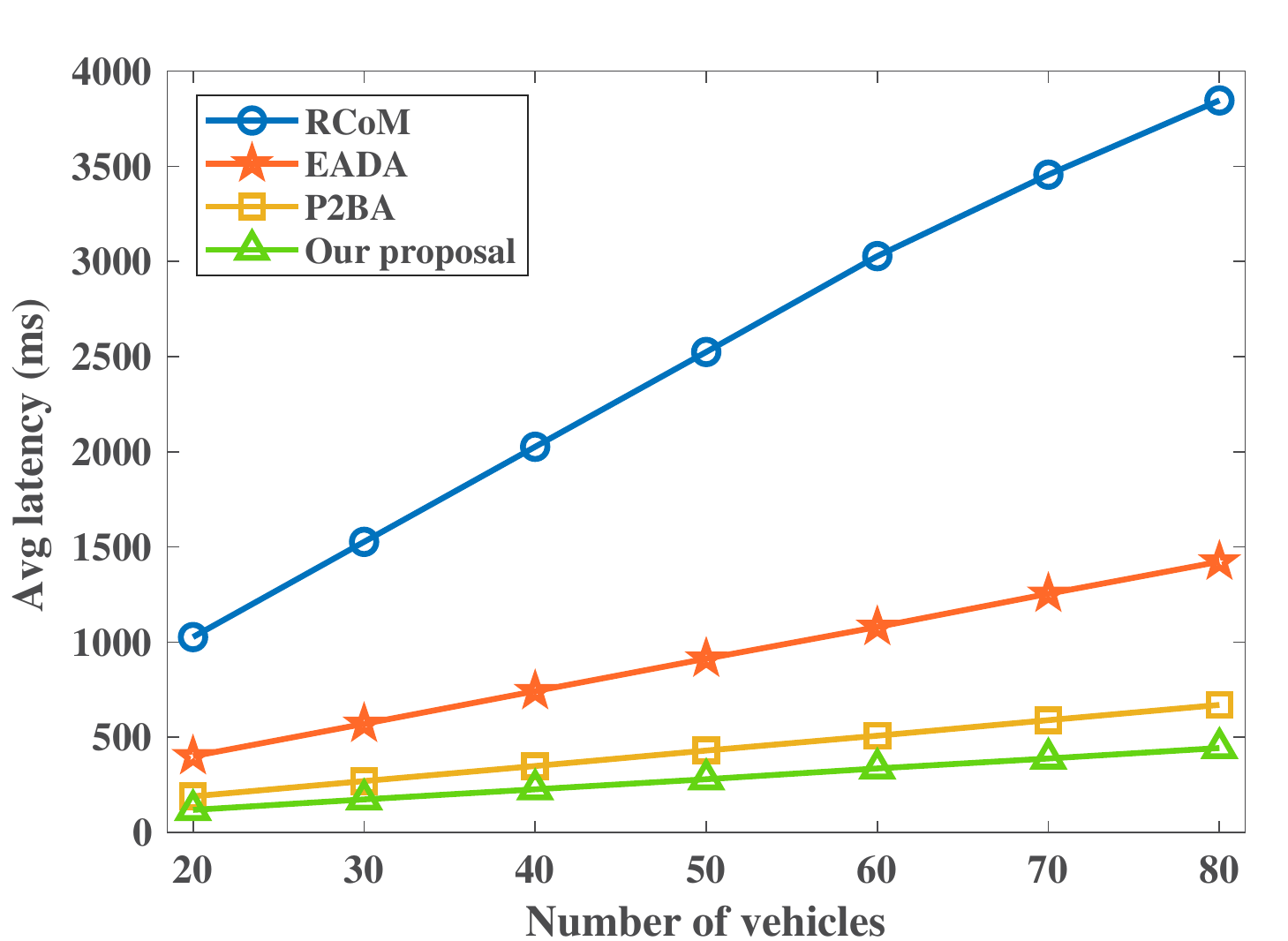}
	\label{fig2}
	
		{\textbf{Figure 9:} Comparison of authentication latency at different vehicle densities.}
\end{figure}

Second, we simulate vehicles driving at various speeds in ns-2.35 and calculate the corresponding authentication latency. We assume several vehicles(e.g., 20) are distributed over different lanes, and the speed of vehicles in each lane is roughly distributed in 5-30 m/s. Figure 8 shows that the latency data did not alter significantly during the simulation process, which is good evidence that speed only slightly affects latency.

Finally, we obtain the curve about the relationship between the average authentication latency and the traffic density through the simulation presented in Figure 9. For various approaches, we can observe that the latency has an upward trend with the increase in vehicle density. However, the curve matching our scheme has the smallest slope, which means the average latency of our proposal increases slowly compared with the other approaches. Besides, our scheme can authenticate 80 vehicles within 450ms in the simulation, which is 33.8$\%$ better than P2BA, 68.8$\%$ better than EADA, 88.5$\%$ better than RCoM, respectively.

\subsection{Message Loss Ratio}
To better evaluate the performance of our scheme, we calculate the message loss rate via simulation. The loss rate is defined as the ratio of the number of messages dropped in the routing layer to the total number of messages sent in communication process. Figure 10 presents the variation of the message loss rate following the number and average speed of vehicles increase. When the vehicle density is 150, the message loss rate of our scheme is controlled within 9.6$\%$, while in PBAS\cite{liu2014message} is about 15$\%$. Accordingly, our approach outperforms those schemes and satisfies the communication requirements in IoV.

\begin{figure}
	\centering
	\includegraphics[width=9cm]{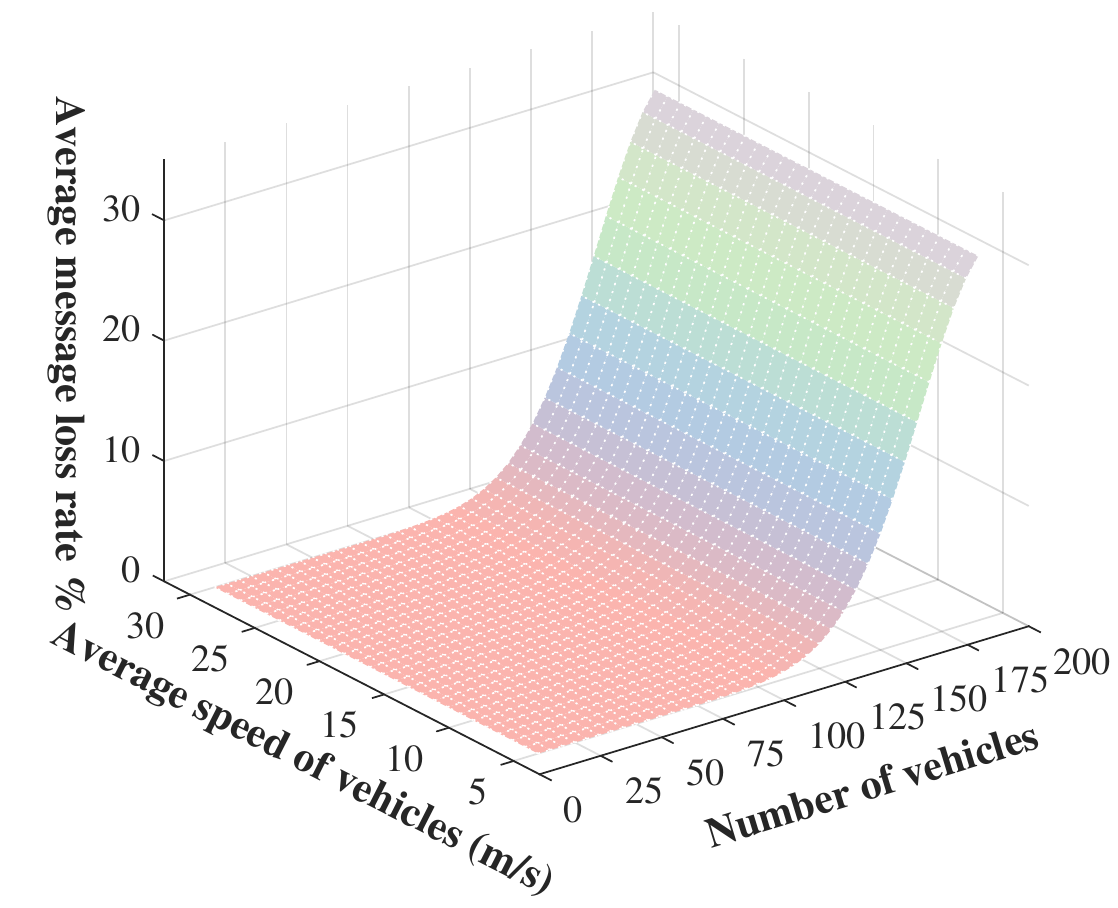}
	\label{fig2}
	
		{\textbf{Figure 10:} Average message loss ratio at different vehicle densities and speeds.}
\end{figure}

\section{Conclusion and Future Work}
Research on addressing the authentication latency issue caused by verifiers having to connect with the blockchain network in advance. We propose a privacy-preserving blockchain-based authentication protocol(PBAG), where RA broadcasts a public global commitment based on all valid certificates. Instead of querying certificates stored in the blockchain, the vehicle can be efficiently proved to be an authorized user by utilizing this global commitment through bilinear pairing. Moreover, our scheme can prevent vehicles equipped invalid certificates from accomplishing the authentication, thus avoiding the time-consuming for checking Certificate Revocation List (CRL).  Finally, we conduct the extra theoretical analysis and simulation to indicate privacy properties and performance of the proposed scheme.  In future work, we plan to implement our scheme in an automated way, and facilitate it to interface with real-world data. Besides, based on the proposed scheme, how to achieve batch authentication in V2V  is also one of our focus.

\appendices
\section{Calculate the derivative value of the function D(X) when X={$\omega^i$}}
let $D(X)=\prod_{i\in [0,n)}(X-\omega^i)= X^n-1$ and $\varphi(X)=D(X)/(X-\omega^i)$, then:
{\setlength\abovedisplayskip{0.2cm}
\setlength\belowdisplayskip{0.2cm}
\begin{flalign}
D^{\prime}(X)&=\sum_{j\in [0,n)}D(X)/(X-\omega^j) & \nonumber \\
&\Rightarrow D^{\prime}(\omega^i)=\prod_{j\in[0,i)}(\omega^i-\omega^j)\cdot\prod_{r\in{(i,n)}}(\omega^i-\omega^r)=\varphi(\omega^i) & 
\end{flalign}}
Thus, rewriting $\varphi(X)$
{\setlength\abovedisplayskip{0.2cm}
\setlength\belowdisplayskip{0.2cm}
\begin{flalign}
\varphi(X) = &\frac{X^n-1}{X-\omega^i}=(\omega^i)^0X^{n-1}+(\omega^i)^1X^{n-2}+(\omega^i)^2X^{n-3} & \nonumber \\
&+\ldots+(\omega^i)^{n-2}X^1+(\omega^i)^{n-1}X^{0}=\sum_{\gamma=0}^{n-1}(\omega^i)^{\gamma}X^{n-1-{\gamma}} &
\end{flalign}}
Finally, evaluating $D^{\prime}(\omega^i)$ at $X=\omega^i$:
{\setlength\abovedisplayskip{0.2cm}
\setlength\belowdisplayskip{0.2cm}
\begin{flalign}
D^\prime(\omega^i) &= \varphi(\omega^i)=(\omega^i)^0\omega^{i(n-1)}+(\omega^i)^1\omega^{i(n-2)} & \nonumber \\
&+(\omega^i)^2\omega^{i(n-3)}+\ldots+(\omega^i)^{n-2}\omega^{(i\cdot 1)}+(\omega^i)^{n-1}\omega^{(i \cdot 0)} & \nonumber \\
&=\sum_{\gamma=0}^{n-1}(\omega^i)^{\gamma}\omega^{i(n-1-{\gamma})}=n\omega^{i(n-1)}=n(\omega^{i\cdot n-i})=n\omega^{-i} &  
\end{flalign}}
Thus, we can compute $D^{\prime}(\omega^i)=n\omega^{-i}$.

\section{Multiple evaluations proving and parameter $c_i$ derivation process}
The first is the multiple evaluation proving process. We have known $c_i=1/D^\prime(\omega^i)$, $\pi_I=\prod_{i\in I}\pi^{c_i}_i$ and $q_I(X) = \frac{\Psi(X)-R(X)}{D_I(X)}$. That is:
{\setlength\abovedisplayskip{0.3cm}
\setlength\belowdisplayskip{0.3cm}
\begin{flalign}
e(C/&g^{R(\tau)},g)=e(\pi_I,g^{D_I(\tau)}) & \nonumber \\
&\Leftrightarrow e(C/g^{R(\tau)},g)=e(\prod_{i\in I}\pi^{c_i}_i,g^{D_I(\tau)}) & \nonumber \\
&\Leftrightarrow e(g, g)^{\Psi(\tau)-R(\tau)}=e(\prod_{i\in I}(g^{q_i(\tau)})^{c_i},g^{D_I(\tau)}) & \nonumber \\
&\Leftrightarrow e(g, g)^{\Psi(\tau)-R(\tau)}=e(g^{\sum_{i\in I}{q_i(\tau)}{c_i}},g^{D_I(\tau)}) &  \\
&\Leftrightarrow e(g, g)^{\Psi(\tau)-R(\tau)}=e(g^{q_{I}(\tau)},g^{D_I(\tau)}) & \nonumber \\
&\Leftrightarrow \Psi(\tau)=q_{I}(\tau) \cdot D_I(\tau)+R(\tau) & \nonumber 
\end{flalign}}
According to the construction of multiple evaluation proving. We will show the process of finding $\pi_I$ that satisfies the conditions of the multiple evaluation proving:
\begin{small}
\begin{flalign}
\begin{split}
q(X)&=\frac{\Psi(X)-R(X)}{D_I(X)}\\
&=\Psi(X)\frac{1}{D_I(X)}-R(X)\frac{1}{D_I(X)}\\
&=\Psi(X)\sum_{i\in I}\frac{1}{D^{\prime}_I(\omega^i)(X-\omega^i)}\\
& \qquad \qquad \qquad -[D_I(X)\sum_{i\in I}\frac{u_i}{D^{\prime}_I(\omega^i)(X-\omega^i)}]\cdot \frac{1}{D_I(X)}\\
&=\sum_{i\in I}\frac{\Psi(X)}{D^{\prime_I}(\omega^i)(X-\omega^i)}-\sum_{i\in I}\frac{u_i}{D^{\prime}_I(\omega^i)(X-\omega^i)}\\
&=\sum_{i\in I}\frac{1}{D^{\prime}_I(\omega^i)}\cdot \frac{\Psi(X)-u_i}{X-\omega^i}\\
&=\sum_{i\in I}\frac{1}{D^{\prime}_I(\omega^i)}\cdot q_i(X)
\end{split}&
\end{flalign}
\end{small}

Thus, we can compute $c_i=1/D^{\prime}_I(\omega^i)$ and $\pi_I=\prod_{i\in I}\pi^{c_i}_i$.

\section{Update of function $\Psi(X)$}
Without loss of generality, we assume the polynomial $\Psi(X)$ is generated for given points $(x_i,y_i)_{i\in [0,n)}$ and satisfies $\Psi(x_i)=y_i$. Based on the Lagrangian interpolation, we can obtain:
\begin{equation}
    \begin{aligned}
\Psi(X) = {\sum}_{i\in[0, n)}\mathcal{L}_i(X)y_i
    \end{aligned}
\end{equation}
where $\mathcal{L}_i(X) = {\prod}_{j\in[0, n), j{\neq }i}\frac{X-x_j}{x_i-x_j}$.

We assume the change in $y_i$ is $\delta$,then:
\begin{small}
\begin{equation}
    \begin{aligned}
\Psi^{\prime}(X)
&={\sum}_{k\in [0,i)}\mathcal{L}_k(X)y_k\\
& \qquad \qquad \quad  +\mathcal{L}_i(x)(y_i+\delta)+{\sum}_{k\in (k,n)}\mathcal{L}_k(X)y_k\\
&={\sum}_{k\in [0,i)}\mathcal{L}_k(X)y_k\\
& \qquad \qquad \quad  +\mathcal{L}_i(x)y_i+\mathcal{L}_i(x)\delta+{\sum}_{k\in (k,n)}\mathcal{L}_k(X)y_k\\
&={\sum}_{k\in [0,n)}\mathcal{L}_k(X)y_k+\mathcal{L}_i(x) \cdot \delta\\
&=\Psi(X)+\mathcal{L}_i(x) \cdot \delta
    \end{aligned}
\end{equation}
\end{small}
Finally, we can obtain the newest $\Psi^{\prime}(X)=\Psi(X)+\delta \cdot \mathcal{L}_i(x)$.

\section{Partial fraction decomposition}
We define \begin{small}$D_I(X)=\displaystyle\prod_{i\in I}(X-x^i)$\end{small}, $I\subset [0,n)$ and rewriting the Lagrange polynomial for interpolating $\Psi(X)$ given all $(\Psi(x_i))_{i\in I}$:

\begin{small}
\begin{equation}
\begin{aligned}
    \mathcal{L}_i(X)= \displaystyle\prod_{j\in I,j\neq i}\frac{X-x^j}{x^i-x^j}=\frac{D_I(X)}{D^{\prime}_I(x^i)(X-x^i)}
\end{aligned}
\end{equation}
\end{small}

where \begin{small} $D_{I}^{\prime}(X)=\sum_{j\in I}D_I(X)/(X-x^j)$\end{small}.

Noting that $D_{I}^{\prime}$ is the derivative of $D_I(X)$\cite{von2013fast}. Therefore, we can rewrite the Lagrange interpolation as:

\begin{small}
\begin{equation}
\Psi_I(X)=\sum_{i\in I}\mathcal{L}_i(X)y_i=D_I(X)\sum_{i\in I}\frac{y_i}{D^{\prime}_I(x^i)(X-x^i)}
\end{equation}
\end{small}
From the properties of Lagrangian polynomials, we know that $\Psi(x_i)=y_i$ for the point $(x_i,y_i)$. Therefore, for the $\Psi(X)=1$, this implies:
\begin{small}
{\setlength\abovedisplayskip{0.2cm}
\setlength\belowdisplayskip{0.2cm}
\begin{flalign}
\Psi_I(X) &= D_I(X)\sum_{i\in I}\frac{y_i}{D^{\prime}_I(x^i)(X-x^i)} & \nonumber \\
 & \xRightarrow{\Psi(X)=1} \frac{1}{D_I(X)} =\sum_{i\in I}\frac{1}{D^{\prime}_I(x^i)(X-x^i)} &
\end{flalign}}
\end{small}
\begin{small}
{\setlength\abovedisplayskip{0cm}
\setlength\belowdisplayskip{0.2cm}
\begin{flalign}
\frac{1}{D_I(X)} &= \frac{1}{\prod_{i\in I}(X-x^i)}=\displaystyle\sum_{i\in I}c_i\cdot\frac{1}{X-x^i}, c_i=\frac{1}{D^{\prime}_{I}(x^i)} & 
\end{flalign}}
\end{small}
we can compute the $D_I(X)$ in $O(|I|log^2|I|)$ time, $D^{\prime}_I(X)$ in $O(|I|)$ time and evaluated at all $x^i$ in $O(|I|log^2|I|)$ time\cite{von2013fast}. Thus, all $c_i$'s can be computed in $O(|I|log^2|I|)$ time. When the $I=[0,n)$, we have:
\begin{small}
{\setlength\abovedisplayskip{0.1cm}
\setlength\belowdisplayskip{0.1cm}
\begin{equation}
    \begin{aligned}
 D_I(X) = \prod_{i\in [0,n)}(X-x^i)= X^n-1\ and\ D^{\prime}(x^i)=nx^{-i}
    \end{aligned}
\end{equation}}
\end{small}
The derivation details of the formula are in Appendix A. Therefore, we can compute any $c_i$ in $O(1)$ time.



\ifCLASSOPTIONcaptionsoff
  \newpage
\fi

\end{document}